\documentclass[apj]{emulateapj}

\usepackage{amsmath,amssymb}
\usepackage{mathptmx}
\usepackage{graphicx}
\usepackage{datetime}

\usepackage[backref,breaklinks,colorlinks,citecolor=blue]{hyperref}

\usepackage[all]{hypcap} 
\def\sectionautorefname~#1\null{Section~#1\null}
\def\subsectionautorefname~#1\null{Section~#1\null}
\def\subsubsectionautorefname~#1\null{Section~#1\null}

\newcommand{\co   }{COSMOS }
\newcommand{\XVP  }{\emph{Chandra COSMOS Legacy} survey}
\newcommand{\Nrawhb}{14} 

\newcommand{\Ndata}{14} 
\newcommand{\Ntot }{14} 
\newcommand{\Nhb  }{10} 
\newcommand{\Nha  }{four}
\newcommand{\Ngood}{12} 

\newcommand{\zrange}{\ifmmode z\simeq3-3.7 \else $z\simeq3-3.7$\fi}
\newcommand{\zpaper}{\ifmmode z\simeq3.3 \else $z\simeq3.3$\fi}
\newcommand{\mysobj}{CID-947}

\newcommand{\ltsim}{\raisebox{-.5ex}{$\;\stackrel{<}{\sim}\;$}}
\newcommand{\gtsim}{\raisebox{-.5ex}{$\;\stackrel{>}{\sim}\;$}}

\newcommand{\todo}{\ifmmode {\Huge \bullet} \else {\Huge$\bullet$}\fi}
\newcommand{\til}{\ifmmode \sim \else $\sim$\fi}

\newcommand{\kms}	{\ifmmode {\rm km\,s}^{-1} \else km\,s$^{-1}$\fi}
\newcommand{\cmii}	{\ifmmode {\rm cm}^{-2}    \else cm$^{-2}$\fi}
\newcommand{\ergs}	{\ifmmode {\rm erg\,s}^{-1} \else erg s$^{-1}$\fi}
\newcommand{\ergcms}	{\ifmmode {\rm erg\,cm}^{-2}\,{\rm s}^{-1} \else erg\,cm$^{-2}$\,s$^{-1}$\fi}
\newcommand{\kev}	{\ifmmode {\rm keV} \else keV\fi}
\newcommand{\mic}	{\ifmmode {\rm \mu m} \else $\mu$m\fi}
\def\arcsec{\hbox{$^{\prime\prime}$}}

\newcommand{\Msun}{\ifmmode M_{\odot} \else $M_{\odot}$\fi}
\newcommand{\Lsun}{\ifmmode L_{\odot} \else $L_{\odot}$\fi}
\newcommand{\mpyr}{\ifmmode \Msun\,{\rm yr}^{-1} \else $\Msun\,{\rm yr}^{-1}$\fi}
\newcommand{\Hubble}	{\ifmmode {\rm km\,s}^{-1}\,{\rm Mpc}^{-1} \else km\,s$^{-1}$\,Mpc$^{-1}$\fi}
\newcommand{\NDunit}	{\ifmmode {\rm Mpc}^{-3} \else Mpc$^{-3}$\fi}
\newcommand{\Msol}{\Msun}

\newcommand{  \Halpha   }{\ifmmode {\rm H}\alpha \else H$\alpha$\fi}
\newcommand{  \halpha   }{\Halpha}
\newcommand{  \ha       }{\Halpha}
\newcommand{  \Hbeta    }{\ifmmode {\rm H}\beta \else H$\beta$\fi}
\newcommand{  \hbeta    }{\Hbeta}
\newcommand{  \hb       }{\Hbeta}
\newcommand{  \CIV      }{\ifmmode {\rm C}\,\textsc{iv}\,\lambda1549 \else C\,\textsc{iv}\,$\lambda1549$\fi}
\newcommand{  \civ      }{\ifmmode {\rm C}\,\textsc{iv}  \else C\,\textsc{iv}\fi}
\newcommand{  \heii     }{\ifmmode {\rm He}\,\textsc{ii} \else He\,\textsc{ii}\fi}
\newcommand{  \HeIIop   }{\ifmmode {\rm He}\,\textsc{ii}\,\lambda4686 \else He\,\textsc{ii}\,$\lambda4686$\fi}
\newcommand{\oiii}{\ifmmode \left[{\rm O}\,\textsc{iii}\right] \else [O\,{\sc iii}]\fi}
\newcommand{\OIII}{\ifmmode \left[{\rm O}\,\textsc{iii}\right]\,\lambda5007 \else [O\,{\sc iii}]\,$\lambda5007$\fi}
\newcommand{  \feii     }{\ifmmode {\rm Fe}\,\textsc{ii} \else Fe\,\textsc{ii}\fi}
\newcommand{  \mgii     }{\ifmmode {\rm Mg}\,\textsc{ii} \else Mg\,\textsc{ii}\fi}
\newcommand{  \MgII     }{\ifmmode {\rm Mg}\,\textsc{ii}\,\lambda2798 \else Mg\,\textsc{ii}\,$\lambda2798$\fi}
\newcommand{  \Lya      }{\ifmmode {\rm Ly}\alpha \else Ly$\alpha$\fi}
\newcommand{  \nv       }{\ifmmode {\rm N}\,\textsc{v}   \else N\,\textsc{v}\fi}

\newcommand{ \fwhm  }{\ifmmode {\rm FWHM} \else FWHM\fi} 
\newcommand{ \fwhb  }{\ifmmode {\rm FWHM}\left(\hb\right) \else FWHM(\hb)\fi}
\newcommand{\sighb  }{\ifmmode \sigma\left(\hb\right) \else $\sigma\left(\hb\right)$\fi}
\newcommand{ \ewhb  }{\ifmmode {\rm EW}\left(\hb\right) \else EW(\hb)\fi}
\newcommand{ \fwha  }{\ifmmode {\rm FWHM}\left(\ha\right) \else FWHM(\ha)\fi}
\newcommand{ \ewha  }{\ifmmode {\rm EW}\left(\ha\right) \else EW(\ha)\fi}
\newcommand{ \lamLlam  }{\ifmmode \lambda L_{\lambda} \else $\lambda L_{\lambda}$\fi}
\newcommand{ \Lop      }{\ifmmode L_{5100} \else $L_{5100}$\fi}
\newcommand{  \Luv      }{\ifmmode L_{1450} \else $L_{1450}$\fi}
\newcommand{ \Lhb   }{\ifmmode L_{\hb} \else $L_{\hb}$\fi}
\newcommand{ \Lha   }{\ifmmode L_{\ha} \else $L_{\ha}$\fi}
\newcommand{\fbol}{\ifmmode f_{\rm bol} \else $f_{\rm bol}$\fi}
\newcommand{\fbolwv}{\ifmmode f_{\rm bol}\left(\lambda\right) \else $f_{\rm bol}\left(\lambda\right)$\fi}
\newcommand{\fbolopt}{\ifmmode f_{\rm bol}\left(5100{\rm \AA}\right) \else $f_{\rm bol}\left(5100{\rm \AA}\right)$\fi}
\newcommand{\fbolthree}{\ifmmode f_{\rm bol}\left(3000{\rm \AA}\right) \else $f_{\rm bol}\left(3000{\rm \AA}\right)$\fi}
\newcommand{  \Lhard    }{\ifmmode L_{\rm 2-10} \else $L_{\rm 2-10}$\fi}
\newcommand{  \Lsix     }{\ifmmode L_{6200} \else $L_{6200}$\fi}
\newcommand{  \lLisx    }{\ifmmode \log\left(\Lop/\ergs\right) \else $\log\left(\Lop/\ergs\right)$\fi}

\newcommand{  \mbh      }{\ifmmode M_{\rm BH} \else $M_{\rm BH}$\fi}
\newcommand{  \lledd    }{\ifmmode L/L_{\rm Edd} \else $L/L_{\rm Edd}$\fi}
\newcommand{  \Lbol     }{\ifmmode L_{\rm bol} \else $L_{\rm bol}$\fi}

\newcommand{\sigBLR}{\ifmmode \sigma_{\mbox{\tiny BLR}} \else $\sigma_{\mbox{\tiny BLR}}$\fi}
\newcommand{\RBLR}{\ifmmode R_{\rm BLR} \else $R_{\rm BLR}$\fi}

\newcommand{  \tgrow     }{\ifmmode t_{\rm growth} \else $t_{\rm growth}$\fi}
\newcommand{  \tUni      }{\ifmmode t_{\rm U} \else $t_{\rm U}$\fi}
\newcommand{  \Mbhdot	}{\ifmmode \dot{M}_{\rm BH} \else $\dot{M}_{\rm BH}$\fi}
\newcommand{  \Maddot	}{\ifmmode \dot{M}_{\rm AD} \else $\dot{M}_{\rm AD}$\fi}
\newcommand{\fobs}{\ifmmode f_{\rm obs} \else $f_{\rm obs}$\fi}

\newcommand{  \as	}{\ifmmode a_{\rm *} 		\else $a_{\rm *}$\fi}
\newcommand{  \re	}{\ifmmode \eta      	\else $\eta$\fi}
\newcommand{  \mseed    }{\ifmmode M_{\rm seed} \else $M_{\rm seed}$\fi}

\newcommand{\ztpf}{$z \simeq 2.4$}
\newcommand{\ztpt}{$z \simeq 3.3$}
\newcommand{\znetprev}{$z\simeq2.4$ and $\simeq3.3$}
\newcommand{\zfpe}{$z \simeq 4.8$}
\newcommand{\zsix}{$z \simeq 6.2$}

\newcommand{\hband}{\textit{H}-band}
\newcommand{\kband}{\textit{K}-band}
\newcommand{\iab}{\ifmmode i_{\rm AB} \else $i_{\rm AB}$\fi}
\newcommand{\kab}{\ifmmode K_{\rm AB} \else $K_{\rm AB}$\fi}

\newcommand{  \chandra }  {{\it Chandra}}
\newcommand{  \xmm     }  {{\it XMM-Newton}}

\begin{document}

\title{
Faint COSMOS AGNs at \MakeLowercase{$z\sim3.3$} - I. Black Hole Properties and Constraints on Early Black Hole Growth
}

\shorttitle{Early Growth of Moderate-Luminosity AGNs at \zpaper}
\shortauthors{B. Trakhtenbrot et al.}

\author{
B.\ Trakhtenbrot\altaffilmark{1,16},
F.\ Civano\altaffilmark{2,3},
C.\ Megan Urry\altaffilmark{2,4,5},
K.\ Schawinski\altaffilmark{1},
S.\ Marchesi\altaffilmark{2,3,6},
M.\ Elvis\altaffilmark{3},
D.\ J.\ Rosario\altaffilmark{7,8},\\
H.\ Suh\altaffilmark{9,3},
J.\ E.\ Mejia-Restrepo\altaffilmark{10},
B.\ D.\ Simmons\altaffilmark{11,12},
A.\ L.\ Faisst\altaffilmark{13,14},
M.\ Onodera\altaffilmark{1,15}
}

\altaffiltext{1}{Institute for Astronomy, Department of Physics, ETH Zurich,\\ Wolfgang-Pauli-Strasse 27, CH-8093 Zurich, Switzerland}
\altaffiltext{2}{Yale Center for Astronomy and Astrophysics,\\ 260 Whitney Ave., New Haven, CT 06520-8121, USA}
\altaffiltext{3}{Harvard-Smithsonian Center for Astrophysics,\\ 60 Garden St., Cambridge, MA 02138, USA}
\altaffiltext{4}{Department of Physics, Yale University,\\ P.O. Box 208120, New Haven, CT 06520-8120, USA}
\altaffiltext{5}{Department of Astronomy, Yale University,\\ P.O. Box 208101, New Haven, CT 06520-8101, USA}
\altaffiltext{6}{INAF--Osservatorio Astronomico di Bologna,\\ via Ranzani 1, I-40127 Bologna, Italy}
\altaffiltext{7}{Max-Planck-Institut f\"{u}r Extraterrestrische Physik (MPE),\\ Postfach 1312, D-85741 Garching, Germany}
\altaffiltext{8}{Department of Physics, Durham University,\\ South Road, Durham DH1 3LE , UK}
\altaffiltext{9}{Institute for Astronomy, University of Hawaii,\\ 2680 Woodlawn Drive, Honolulu, HI 96822, USA}
\altaffiltext{10}{Departamento de Astronom\'{\i}a, Universidad de Chile,\\ Camino el Observatorio 1515, Santiago, Chile}
\altaffiltext{11}{Oxford Astrophysics,\\ Denys Wilkinson Building, Keble Road, Oxford OX1 3RH, UK}
\altaffiltext{11}{Oxford Astrophysics,\\ Denys Wilkinson Building, Keble Road, Oxford OX1 3RH, UK}
\altaffiltext{13}{Infrared Processing and Analysis Center, California Institute of Technology, Pasadena, CA 91125, USA}
\altaffiltext{14}{Cahill Center for Astronomy and Astrophysics, California Institute of Technology, Pasadena, CA 91125, USA}
\altaffiltext{15}{Subaru Telescope, National Astronomical Observatory of Japan, 650 North A'ohoku Place, Hilo, HI 96720, USA}
\altaffiltext{16}{Zwicky Postdoctoral Fellow}
\email{benny.trakhtenbrot@phys.ethz.ch}

\slugcomment{Draft Version: \currenttime{} on \today}

\begin{abstract}
We present new Keck/MOSFIRE $K$-band spectroscopy for a sample of \Ntot\ faint, X-ray-selected active galactic nuclei (AGNs) in the COSMOS field. 
The data cover the spectral region surrounding the broad Balmer emission lines, which enables the estimation of black hole masses (\mbh) and accretion rates (in terms of \lledd).
We focus on 10 AGNs at \ztpt, where we observe the \hbeta\ spectral region, while for the other four \ztpf\ sources we use the \Halpha\ broad emission line.
Compared with previous detailed studies of unobscured AGNs at these high redshifts, our sources are fainter by an order of magnitude, corresponding to number densities of order $\sim10^{-6}-10^{-5}\,\NDunit$. 
The lower AGN luminosities also allow for a robust identification of the host galaxy emission, necessary to obtain reliable intrinsic AGNs luminosities, BH masses and accretion rates.
We find the AGNs in our sample to be powered by supermassive black holes (SMBHs) with a typical mass of $\mbh\simeq5\times10^{8}\,\Msol$ -- significantly lower than the higher-luminosity, rarer quasars reported in earlier studies. 
The accretion rates are in the range $\lledd\sim0.1-0.4$, with an evident lack of sources with lower \lledd\ (and higher \mbh), as found in several studies of faint AGNs at intermediate redshifts.
Based on the early growth expected for the SMBHs in our sample, we argue that a significant population of faint $z\sim5-6$ AGNs, with $\mbh\sim10^{6}\,\Msol$, should be detectable in the deepest X-ray surveys available, but this is \emph{not} observed. 
We discuss several possible explanations for the apparent absence of such a population, concluding that the most probable scenario involves an evolution in source obscuration and/or radiative efficiencies.
\end{abstract}

\section{Introduction}
\label{sec:intro}

While the local Universe provides ample evidence for the existence of supermassive black holes (SMBHs) with masses of $\mbh\sim10^6-10^{10}\,\Msol$ in the centers of most galaxies \cite[e.g.,][and references therein]{KormendyHo2013_MM_Rev}, the understanding of their growth history relies on the analysis of accreting SMBHs, observed as active galactic nuclei (AGNs).
Several studies and lines of evidence, mainly based on the observed redshift-resolved luminosity functions of AGNs, suggest that the epoch of peak SMBH growth occurred at $z\sim2-3$, in particular in the sense of a peak in the integrated accretion density \cite[e.g.,][and references therein]{Marconi2004,Hasinger2005,Ueda2014,Aird2015_XLF,BrandtAlexander2015_Rev}.
Recent results from increasingly deep surveys have shown that at yet higher redshifts the number density and integrated emissivity of AGNs experience a marked decrease \cite[e.g.,][]{Brusa2009_XCOS_hiz,Civano2011_hiz,McGreer2013_QLF_z5,Vito2014_XLF,Miyaji2015_XLF}.
Phenomenological ``synthesis models'' have been used to account for the observed evolution of the AGN population out to $z\sim4-5$, particularly based on deep X-ray surveys \cite[see, e.g.,][]{Gilli2007,Treister2009_XBR,Ueda2014,Aird2015_XLF,Georgakakis2015_XLF_hiz}.
Broadly speaking, these synthesis models successfully reproduce the population of relic SMBHs in the local Universe, the X-ray background radiation, and the X-ray number counts.
However, all these models depend on several simplifying assumptions, including the accretion rates, radiative efficiencies, and the shape of the X-ray spectral energy distribution (SED) of AGNs, among others. 
Our current understanding of the early growth of SMBHs is therefore still extremely limited. 
Most importantly, it lacks robust characterization of the distributions of the most basic physical properties of accreting SMBHs: black hole masses (\mbh), accretion rates (in terms of \lledd\ or \Mbhdot) and radiative efficiencies (\re; and/or BH spins, \as), for SMBHs across a wide range of activity phases.

Reliable estimates of \mbh, and therefore \lledd, from single-epoch spectra of AGNs at considerable redshifts rely on the careful analysis of either the spectral regions surrounding either the \Hbeta, \Halpha, or \MgII\ broad emission lines, and on the results of reverberation mapping campaigns.
Other emission lines, which may potentially enable the estimation of \mbh\ in tens of thousands of quasars from the Sloan Digital Sky Survey (SDSS), up to $z\sim5$ (e.g., \CIV), are known to be problematic \cite[e.g.,][]{Baskin2005,Shen_dr5_cat_2008, Fine2010_CIV,Shen_Liu_2012,TrakhtNetzer2012_Mg2,Tilton2013_COS_CIV}.
Therefore, at $z\gtsim2$, the study of the evolution of \mbh\ practically requires near-IR (NIR) spectroscopy, and ground-based studies are thus limited to specific redshift bands, 
at $z\sim 2.1-2.7$, $3.1-3.8$, $4.6-4.9$, and $6-7.2$.
Several studies followed this approach with relatively small samples of optically selected, high-luminosity unobscured AGNs, mostly focusing on the most luminous sources at each redshift bin \cite[e.g.,][]{Shemmer2004,Kurk2007,Netzer2007_MBH,Dietrich2009_Hb_z2,Marziani2009,Willott2010_MBH,DeRosa2011,Trakhtenbrot2011}.
The studies of \cite{Shemmer2004} and \cite{Netzer2007_MBH} clearly show that the most massive BHs in the Universe, with $\mbh\gtrsim10^{10}\,\Msol$ \cite[][]{McConnell2011_MBH_10} are already in place by $z\sim3.5$, powering some of the most luminous quasars at $z\sim3-4$.
Given their extreme masses, but modest accretion rates of $\lledd\simeq0.2$, these objects must have grown at higher rates at yet earlier epochs.
Indeed, a population of SMBHs with $\mbh\sim10^{9}\,\Msun$ is now well established at $5 \lesssim z \lesssim7$, presenting rapid, Eddington-limited accretion  \cite[e.g.,][]{Kurk2007,Willott2010_MBH,DeRosa2011,Trakhtenbrot2011,DeRosa2014}.
Thus, the extremely luminous $z\sim3-4$ quasars studied to date mark the final stage of the early, rapid growth of the most massive BHs in the Universe.

These results motivated the development of new models for the formation of high-mass BH seeds, at $z\gtsim10$.
Such processes, involving either dense stellar environments or direct collapse of gaseous halos, may lead to BH seeds with masses of up to $\mseed\sim10^4$ or $10^6\,\Msol$, respectively \cite[see reviews by][and references therein]{Volonteri2010_rev,Natarajan2011_seeds_rev}.
Some models predict that such massive BH seeds are sufficiently abundant in the early Universe to easily account for the rare $\mbh\sim10^9\,\Msol$ quasars observed at $z>3$ \cite[see, e.g.,][]{Dijkstra2008_seeds_ND,Agarwal2013_obese}.
Several other recent studies have instead focused on extremely efficient accretion onto seed BHs, as an alternative (or complementary) explanation for the highest-redshift quasars \cite[e.g.,][]{Alexander2014_seeds,Madau2014_supEdd}.
It is possible that these rare, extremely luminous and massive quasars have indeed grown from high-mass BH seeds and/or by extreme accretion scenarios, while the majority of high-redshift SMBHs, detected as lower-luminosity AGNs, can be explained by stellar remnants, with $\mseed\lesssim100\,\Msol$. 
The only way to observationally test these scenarios and seeding models would be to constrain the distributions of \mbh\ (and \lledd) in large samples of AGNs, which extend toward low luminosities and thus significant number densities.
Moreover, these distributions should be established at the highest possible redshifts, since at later epochs the initial conditions of BH seed formation are completely ``washed out,'' partially due to the increasing importance of ``late seeding'' \cite[e.g.,][]{Schawinski2011_clumpy_seeding,Bonoli2014_late_seeding}.
Such distributions would in turn enable the direct study of the progenitors of the typical luminous SDSS $z\sim1-2$ quasars, which have already accumulated most of their final mass.

Since wide optical surveys (e.g., SDSS) only probe the rarest, most luminous (and least obscured) sources at $z>2$, they cannot provide the parent samples required for mapping the distributions of \mbh\ and \lledd.
The most up-to-date determinations of the AGN luminosity function at these high redshifts indicate that the most luminous quasars have number densities of order $\Phi\sim10^{-8}\,\NDunit$, while AGNs that are fainter by an order of magnitude are more abundant by at least a factor of $20$ \cite[e.g.][]{Glikman2010_QLF,Ikeda2011,Masters2012,McGreer2013_QLF_z5}.
The best sources for samples of these fainter AGNs are deep, multi-wavelength surveys, with appropriate X-ray coverage, such as the COSMOS and CDF-S surveys (\citealt{Civano2016_XVP_cat} and \citealt{Xue2011_CDFS_4Ms}, respectively; see \citealt{BrandtAlexander2015_Rev} for a recent review).
In such surveys, moderate-luminosity AGNs ($L_{\rm X}\gtrsim few\times10^{43}\,\ergs$) can be detected at redshifts as high as $z\sim5$, as confirmed by spectroscopic follow-up campaigns \cite[e.g.,][M15 hereafter]{Szokoly2004,Trump2009a_spec,Silverman2010_ECDFS,Civano2011_hiz,Vito2013_CDFs_hiz,Marchesi2015_XVP_hiz}. 
Furthermore, the multi-wavelength data available in these deep fields can provide a large suite of ancillary information relevant to the evolution of the central accreting SMBHs, ranging from the accretion process and the central engine (i.e., X-ray spectral analysis) to the properties of the host galaxies (e.g., the masses and sizes of the stellar components and/or the presence of cold gas).

We therefore initiated a dedicated project to measure BH masses, accretion rates, and host galaxy properties in a sample of moderate-luminosity, $z\sim2.1-3.7$ AGNs, located within the COSMOS field \cite[][]{Scoville2007_COSMOS_overview}, and selected through the extensive X-ray coverage provided by the relevant \chandra\ surveys \cite[][]{Elvis2009_CC1,Civano2016_XVP_cat}.
In this paper we present new Keck/MOSFIRE NIR spectroscopy and determinations of \mbh\ and \lledd\ for a sample of \Ntot\ such objects. 
In \autoref{sec:obs_data_analysis} we describe the observations, data reduction, and analysis, including the estimates of \mbh\ and \lledd. 
In \autoref{sec:res_and_disc} we compare these, and other probes of SMBH evolution, to those found for more luminous quasars, and examine the relevance of high-mass BH seeding models to lower-luminosity AGNs.
We summarize the main findings of this study in \autoref{sec:summary}.
We note that one particularly intriguing object in our sample (\mysobj) was discussed extensively in a previous, separate publication \cite[][T15 hereafter]{Trakhtenbrot2015_CID947}.
Throughout this work we assume a cosmological model with $\Omega_{\Lambda}=0.7$, $\Omega_{\rm M}=0.3$, and
$H_{0}=70\,\kms\,{\rm Mpc}^{-1}$.

%
\section{Sample, Observations, and Data Analysis}
\label{sec:obs_data_analysis}

\subsection{Sample Selection and Properties}
\label{subsec:sample}

This study focuses on \Ndata\ AGNs, selected from the X-ray \emph{Chandra} catalog of the COSMOS field.
The \chandra\ data combine the \chandra-COSMOS survey \cite[][]{Elvis2009_CC1,Civano2012_CC3}, and the more recent \XVP\ \cite[][]{Civano2016_XVP_cat,Marchesi2016_XVP_opt_ID}.
We note that all \Ndata\ sources are also detected in the \xmm\ X-ray survey of the COSMOS field \cite[][see below]{Hasinger2007_COSMOS_XMM}.
We selected sources that are robustly classified as broad-line AGNs at \zrange, based on the (optical) spectroscopic surveys of the COSMOS field \cite[][]{Lilly2007_zCOSMOS,Lilly2009,Trump2009a_spec}.
The chosen redshift range ensures that the spectral region surrounding the \hb\ broad emission line will be observed in the $K$-band.
Adequate coverage of this spectral region is essential for the estimation of \mbh\ \cite[e.g.,][]{TrakhtNetzer2012_Mg2,Shen2013_rev}.
All the sources are robustly detected in the \kband, based on the UltraVISTA DR2 catalog \cite[see survey description in][]{McCracken2012_COSMOS_UltraVISTA}. 
To ensure an adequate signal-to-noise ratio (S/N) within a reasonable observation time, we further focused on those \zrange\ COSMOS AGNs that meet a flux limit of $\kab\leq21.5$, resulting in \Nrawhb\ targets in the range $20<\kab<21.5$.
Four additional broad-line AGNs at \ztpf\ were selected to be observed in parallel to (some) of the primary targets, within the same MOSFIRE masks. 
For these \Nha\ sources, the $K$-band covers the \Halpha\ broad emission line, which can also be used for \mbh\ estimates \cite[through secondary calibration; see, e.g.,][]{Greene_Ho_Ha_2005}.
These slightly brighter sources ($19.2<\kab<20.1$) were drawn from a much larger population of X-ray-selected, unobscured COSMOS AGNs at this redshift band, solely based on their (angular) proximity to the primary \ztpt\ AGNs. As such, they do \emph{not} represent the general population of \ztpf\ AGNs.
The UltraVISTA \kband\ fluxes are further used here to test the absolute flux calibration of the MOSFIRE spectra (see \autoref{subsec:obs_red} below).

The (full band) X-ray fluxes of the sources span about a factor of 15, $f_{\rm \left[0.5-10\, \kev\right]}\sim \left(2.2-32\right)\times10^{-15}\,\ergcms $, corresponding to rest-frame hard-band luminosities of $L_{\rm \left[2-10\,\kev\right]}\sim \left(7.3-97\right)\times10^{43}\,\ergs$, as reported in M15.
These X-ray fluxes are high enough to allow for a robust detection of \emph{all} of our sources in the \emph{XMM}-COSMOS survey \cite[][]{Hasinger2007_COSMOS_XMM,Brusa2010_COSMOS_XMM}. 
We compare the \chandra- and \emph{XMM}-based X-ray luminosity measurements in \autoref{subsec:ancillary_data} below.
Basic information regarding the sources and the observations (detailed below) is provided in \autoref{tab:obs_log}.
The $z=3.328$ AGN \mysobj\ was analyzed and published separately in T15, because it exhibits several intriguing features, including an extremely high BH mass, extremely low accretion rate, and an AGN-driven outflow, among others.
In many parts of the present study we will mention \mysobj\ separately, as its properties differ from the rest of our sample.

The \kband\ magnitudes of our sources can be used to estimate a lower limit to the BH masses and accretion rates we might expect to find, using the methods detailed in \autoref{subsec:est_mbh_lledd_tg}. 
At $z=3.3$, the chosen flux limit ($\kab\simeq21.5$) translates to $\Lop\simeq1.1\times10^{45}\,\ergs$ and $\Lbol\simeq7\times10^{45}\,\ergs$, assuming the composite quasar spectrum of \cite{VandenBerk2001}, and the \Lop-dependent bolometric correction introduced in \citet[see \autoref{subsec:est_mbh_lledd_tg}]{TrakhtNetzer2012_Mg2}. 
Further assuming that the width of the \hbeta\ line is in the range $1500<\fwhb<15000\,\kms$, we obtain lower limits of $\mbh\gtrsim5.5\times10^{7}\,\Msol$ and of $\lledd\gtrsim0.008$.
\footnote{Note that these limits are strongly (anti-)correlated, i.e., sources with $\mbh\sim6\times10^{7}\,\Msol$ and $\lledd\sim0.01$ would be significantly fainter than our chosen flux limit. See \autoref{fig:LLedd_vs_MBH}.}

\begin{figure}
\centering
\includegraphics[angle=-90, width=0.45\textwidth]{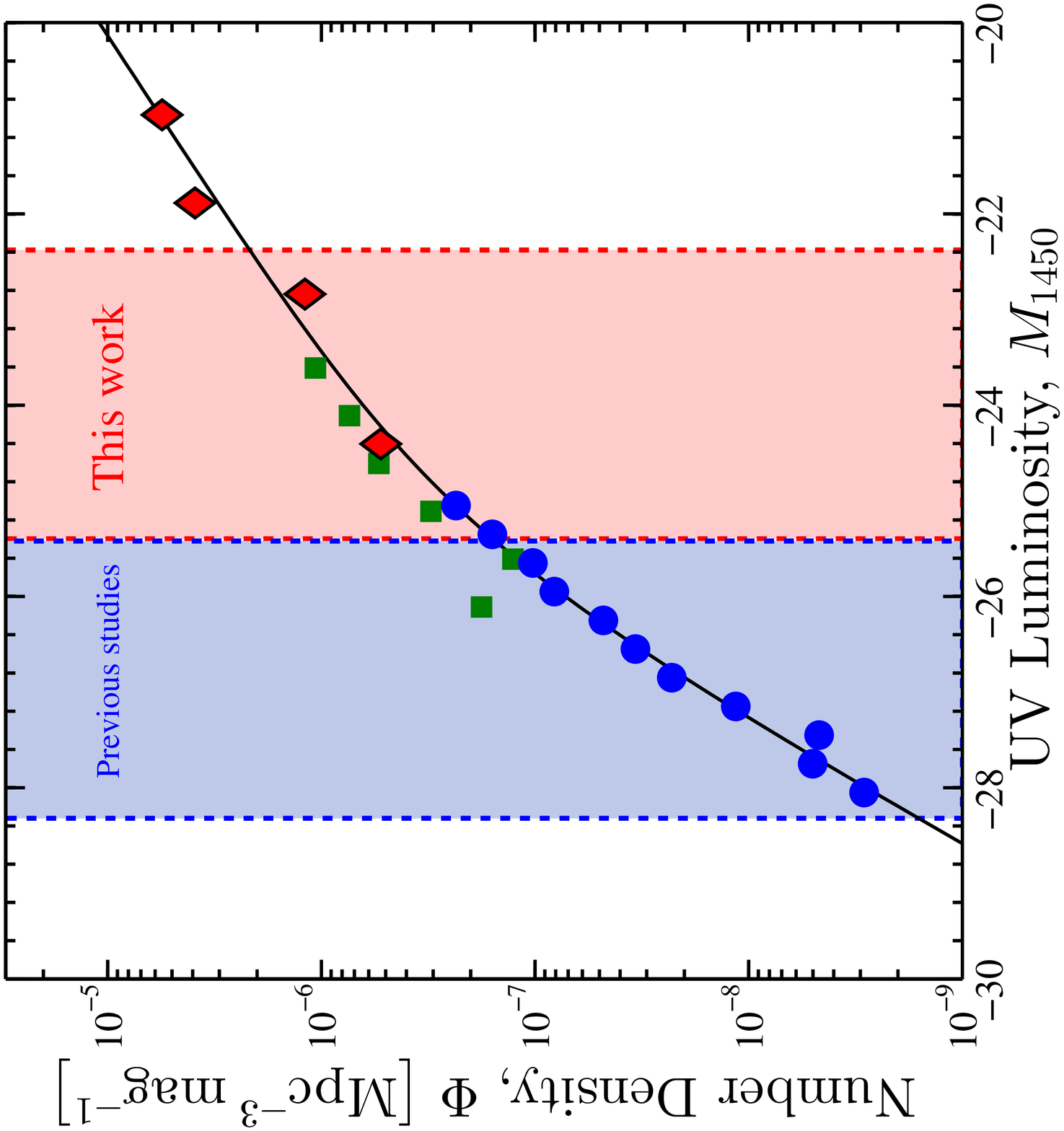} 
\caption{
The luminosity function of unobscured AGNs at $z\sim3-3.5$, reproduced from the study of \cite{Masters2012}, including the best-fit double power-law model (black line).
The red diamonds represent COSMOS AGNs, similar to the parent sample from which our targets are drawn.
Blue circles at higher luminosities are taken from the SDSS \cite[][]{Richards2006_QLF}, while the green squares in the overlap region are taken from the SWIRE survey \cite[][]{Siana2008_QLF_SWIRE}. 
Other samples and error bars from all data points are omitted for clarity.
The shaded regions represent the luminosity regimes covered by our sample (red) and previous studies of \mbh\ and \lledd\ in luminous $z\sim3-4$ AGNs \cite[blue;][]{Shemmer2004,Netzer2007_MBH}.
Our sample probes a much more representative population of \ztpt\ AGNs, with an integrated number density that is higher by a factor of about 25 than the previously studied objects.
}
\label{fig:QLF}
\end{figure}

Compared to previous studies of \mbh\ and \lledd\ in $z\sim3-4$ AGNs \cite[][]{Shemmer2004,Netzer2007_MBH}, our sample covers lower luminosities.
The rest-frame UV luminosities of our \ztpt\ sources, measured from the optical spectra, are in the range $\Luv=\left(0.8-13\right)\times10^{45}\,\ergs$ ($M_{1450}=-25.4$ to $-22.4$; see \autoref{tab:lums}).
The typical UV luminosities are fainter, by about a factor of 6, than those probed in previous studies.
Our sample therefore represents a much more abundant AGN population. 
In \autoref{fig:QLF} we present the luminosity function of unobscured, $z\sim3.2$ AGNs determined by \cite{Masters2012}, which relies on COSMOS AGNs similar to the parent sample of our sources.
The luminosity regimes probed by our sample, and the previously studied samples, are marked.
The integrated number density of sources within the luminosity range we target is $\Phi\simeq2.5\times10^{-6}\,\NDunit$, higher by a factor of about 25 than that of the more luminous, previously studied objects (for which $\Phi\simeq10^{-7}\,\NDunit$).

As our sample is defined through a combination of several criteria, it is worth bearing in mind the possible selection biases.
First, the \chandra-based X-ray selection should include all Compton-thin AGNs above the survey flux limit (i.e., $N_{\rm H} \ltsim 10^{23}\,\cmii$; see M15).
Several studies have highlighted the presence of obscured AGN emission in high-redshift sources \cite[e.g.,][]{Fiore2008,Treister2009}.
Next, the X-ray AGNs must be associated with an optical and NIR counterpart, and have optical spectroscopy for redshift determination and classification as broad-line AGNs.
In principle, this would mean that dust-rich (but Compton-thin) systems, such as ``red quasars'' \cite[e.g.,][]{Banerji2015_redQSOs_hiz,Glikman2015_red_mergers}, may be missed by our sample selection criteria.
However, the M15 compilation of high-$z$ AGNs in the \XVP\ notes that only about 40 X-ray sources among the 4016 X-ray-selected sources ($\sim1\%$) in the entire survey lacked $i$-band counterparts, with about half of those lacking also \kband\ counterparts. 
For most COSMOS AGNs the spectral information is based on the zCOSMOS-bright survey (see \autoref{tab:obs_log}), which imposes an optical flux limit of $i_{\rm AB}<22.5$ \cite[][]{Lilly2007_zCOSMOS}.
Additional spectroscopic follow-up available for COSMOS (X-ray) AGNs provides several other $z>3$ X-ray-selected broad-line sources with $i_{\rm AB}<25$, well beyond the zCOSMOS limit.
These are, however, generally too faint to be included in our sample in terms of the \kband\ cut we imposed, motivated by the need to observe a sizable sample. 
We conclude that our sample is highly representative of the population of unobscured COSMOS AGNs at \zrange, down to a flux limit of $\kab\simeq21.5$.
As the sample is mainly limited by (rest-frame) UV and optical flux selection, it may only be biased against highly obscured AGNs, either in the X-rays or in the (rest-frame) UV. 
Such ``missed'' AGNs may be indeed powered by SMBHs with \mbh\ and/or \lledd\ that are higher than the aforementioned lower limits.
%

\capstartfalse
\begin{deluxetable*}{llcccclccccc}
\tablecolumns{12}
\tablewidth{0pt}
\tablecaption{Observation Log \label{tab:obs_log}}
\tablehead{
\colhead{Subsample} &
\multicolumn{2}{c}{Object ID} &
\colhead{R.A.} &
\colhead{Decl.} &
\colhead{$z$\tablenotemark{c}} &
\colhead{Optical} &
\multicolumn{2}{c}{$K_{\rm AB}$ mag. \tablenotemark{d}} & 
\colhead{MOSFIRE Exp.} &
\colhead{S/N\tablenotemark{e}} &
\colhead{Comments} \\
 & X-ray\tablenotemark{a} & Galaxy ID\tablenotemark{b} & (deg) & (deg) &  & ~~Spec.\tablenotemark{c} & (ref.) & (syn.) & Time (s) &
}
\startdata
\ztpt\  & CID-349 & 1294973 & 150.004380 & 2.038898 & 3.5150 & zCOSb  & 21.238 & 21.277 & 9000  & 7  & ... \\
~~~     & CID-413 & 2039436 & 149.869660 & 2.294046 & 3.3450 & zCOSb  & 20.134 & 20.472 & 5400  & 9  & ... \\
~~~     & CID-113 & 2350265 & 150.208850 & 2.481935 & 3.3330 & zCOSb  & 19.555 & 19.774 & 6840  & 16 & ... \\
~~~     & CID-947 & 1593500 & 150.297250 & 2.148846 & 3.3280 & zCOSb  & 20.052 & 20.045 & 3600  & 8  & ... \\
~~~     & LID-775 & 3176366 & 149.472888 & 2.793379 & 3.6260 & IMACS  & 21.488 & 21.442 & 14400 & 6  & ... \\
~~~     & LID-1638& 1462117 & 150.735585 & 2.199557 & 3.5030 & VVDS   & 19.651 & 19.736 & 3600  & 15 & ... \\
~~~     & LID-205 & 2665989 & 150.240801 & 2.659037 & 3.3560 & zCOSb  & 21.197 & 21.245 & 10800 & 8  & ... \\
~~~     & LID-499 & 2534376 & 150.737172 & 2.722557 & 3.3020 & zCOSb  & 20.215 & 20.378 & 2520  & 6  & ... \\
~~~     & LID-460 & 2583306 & 150.620069 & 2.671382 & 3.1430 & zCOSb  & 19.865 & 20.356 & 4860  & 22 & ... \\
~~~     & LID-721 & 2137194 & 149.529103 & 2.380143 & 3.1080 & IMACS  & 20.157 & 20.010 & 3600  & 9  & ... \\
\hline \\ [-1.75ex]
\ztpf\  & LID-496 & 2577949 & 150.720703 & 2.693635 & 2.6298 & SDSS   & 20.116 & 20.225 & 2520  & 6  & with LID-499 \\
~~~     & LID-504 & 2530857 & 150.767390 & 2.739021 & 2.2220 & zCOS   & 19.670 & 19.736 & 2520  & 8  & with LID-499 \\
~~~     & LID-451 & 2592676 & 150.705563 & 2.629612 & 2.1225 & SDSS   & 19.178 & 19.238 & 4860  & 17 & with LID-460 \\
~~~     & CID-352 & 1300441 & 150.058920 & 2.015179 & 2.4980 & SDSS   & 19.201 & 19.369 & 9000  & 22 & with CID-349
\enddata
\tablenotetext{a}{X-ray object IDs correspond to either the C-\co\ (``CID'') or \XVP\ catalogs (``LID'') (\citealt{Elvis2009_CC1} and \citealt{Civano2016_XVP_cat}, respectively).}
\tablenotetext{b}{COSMOS galaxy IDs correspond to those given by \cite{Capak2007}.}
\tablenotetext{c}{Redshifts are obtained from rest-frame UV emission lines, observed through optical spectroscopy, from either the zCOSMOS-bright \cite[``zCOSb'';][]{Lilly2007_zCOSMOS}, IMACS \cite[][]{Trump2009a_spec}, VVDS \cite[][]{LeFevre2013_VVDS}, or SDSS \cite[DR7;][]{Abazajian2009_DR7} observations of the COSMOS field.}
\tablenotetext{d}{$K$-band magnitudes obtained from the UltraVISTA survey \cite[``ref.'' column,][]{McCracken2012_COSMOS_UltraVISTA} and from synthetic photometry of the calibrated MOSFIRE spectra.}
\tablenotetext{e}{Median signal-to-noise ratios, calculated per spectral bin of 1 \AA\ in the rest frame ($\sim45-60\,\kms$).}
\end{deluxetable*}
\capstarttrue

\subsection{Observations and Data Reduction}
\label{subsec:obs_red}

The Keck/MOSFIRE \cite[][]{McLean2012_MOSFIRE} observations were allocated through the Yale-Caltech collaborative agreement, and conducted during six nights in the period between 2014 January and 2015 February.
Observational conditions during 5 of the nights were generally good, with typical seeing of $\sim$1\arcsec\ (or $\sim$0\farcs8 in the NIR), but also with some periods of high humidity and cloud cover. One night was completely lost due to poor weather.
Our campaign targeted all the \Nrawhb\ primary \ztpt\ targets we selected, except for one source (LID-283).
The targets were observed as part of 12 different MOSFIRE masks, with the four secondary \ztpf\ sources being observed within (some of) the masks designed to include the primary \ztpt\ ones.
To ensure adequate coverage of the sky background emission, and its subtraction from the AGN signal, the sources studied here were observed through two or three MOSFIRE (pseudo-)slits, corresponding to 14\arcsec\ or 21\arcsec, respectively.  
We set the slits to have widths of 0\farcs7-1\arcsec, depending on the seeing. 
This translates to a spectral resolution of $\sim2500-3600$ ($80-120\,\kms$), which is adequate for studies of broad and narrow emission lines in unobscured AGNs.
The rest of the slits in the MOSFIRE masks were allocated to a wide variety of other COSMOS targets, totaling 225 targets and including many X-ray-selected AGNs that lack redshift determinations. 
Those data will be analyzed and published separately.
We also observed several A0V stars (HIP34111, HIP43018, HIP56736, and HIP64248) as well as the fainter white dwarf GD71, at least twice during each night to allow robust flux calibration. 

We reduced the data using a combination of different tools. 
First, we used the dedicated MOSFIRE pipeline\footnote{Version 1.1, released 2015 January 6. See:\\ \url{http://github.com/Keck-DataReductionPipelines/MosfireDRP}} to obtain flat-fielded, wavelength-calibrated 2D spectra of all the sources observed within each mask (including the standard stars). 
The wavelength calibration was performed using sky emission lines, and the best-fit solutions achieved a typical rms of $\sim$0.1 \AA.
Next, we used standard {\tt IRAF} procedures to produce 1D spectra, using apertures in the range of 4-6 pix (i.e., $0\farcs72-1\farcs1$).
Finally, we used the {\tt Spextool} IDL package to remove the telluric absorption features near 2 \mic\ and to perform the relative and absolute flux calibrations, based on a detailed spectrum of Vega \cite[][]{Vacca2003,Cushing2004}.
We verified that the resulting spectra do not have any significant residual spectral features, which might have been misinterpreted as real, AGN-related emission or absorption features.

To test the reliability of our flux calibration procedure, we have calculated the synthetic magnitudes of the calibrated spectra (using the UltraVISTA $K$-band filter curve).
The synthetic magnitudes are generally in good agreement with the reference UltraVISTA magnitudes, with differences of less than 0.2 mag for 11 of the 14 sources in the final sample.
The remaining three sources have flux differences of less than 0.5 mag.
Such differences can be explained by intrinsic AGN variability, which for the roughly year-long timescales probed here is expected to be $\sim$0.2-0.5 mag \cite[e.g.,][]{VandenBerk2004,Wilhite2008,Morganson2014_PS1}. 
We do, however, note that our calibrated spectra are \emph{systematically} fainter than the reference imaging-based fluxes, by about 0.1 mag.
In any case, since $\mbh \sim L^{0.65}$ and $\lledd \sim L^{0.35}$ (see \autoref{subsec:est_mbh_lledd_tg}), these flux differences correspond to uncertainties of less than $\sim$0.1 dex, and most probably $\sim$0.05 dex, on the estimated basic physical properties of the SMBHs under study. 
This is much smaller than the systematic uncertainty associated with the ``virial'' \mbh\ estimator used here \cite[see][and \autoref{subsec:est_mbh_lledd_tg}]{Shen2013_rev}

\begin{figure*}[t]
\centering
\includegraphics[angle=-90,width=1.0\textwidth]{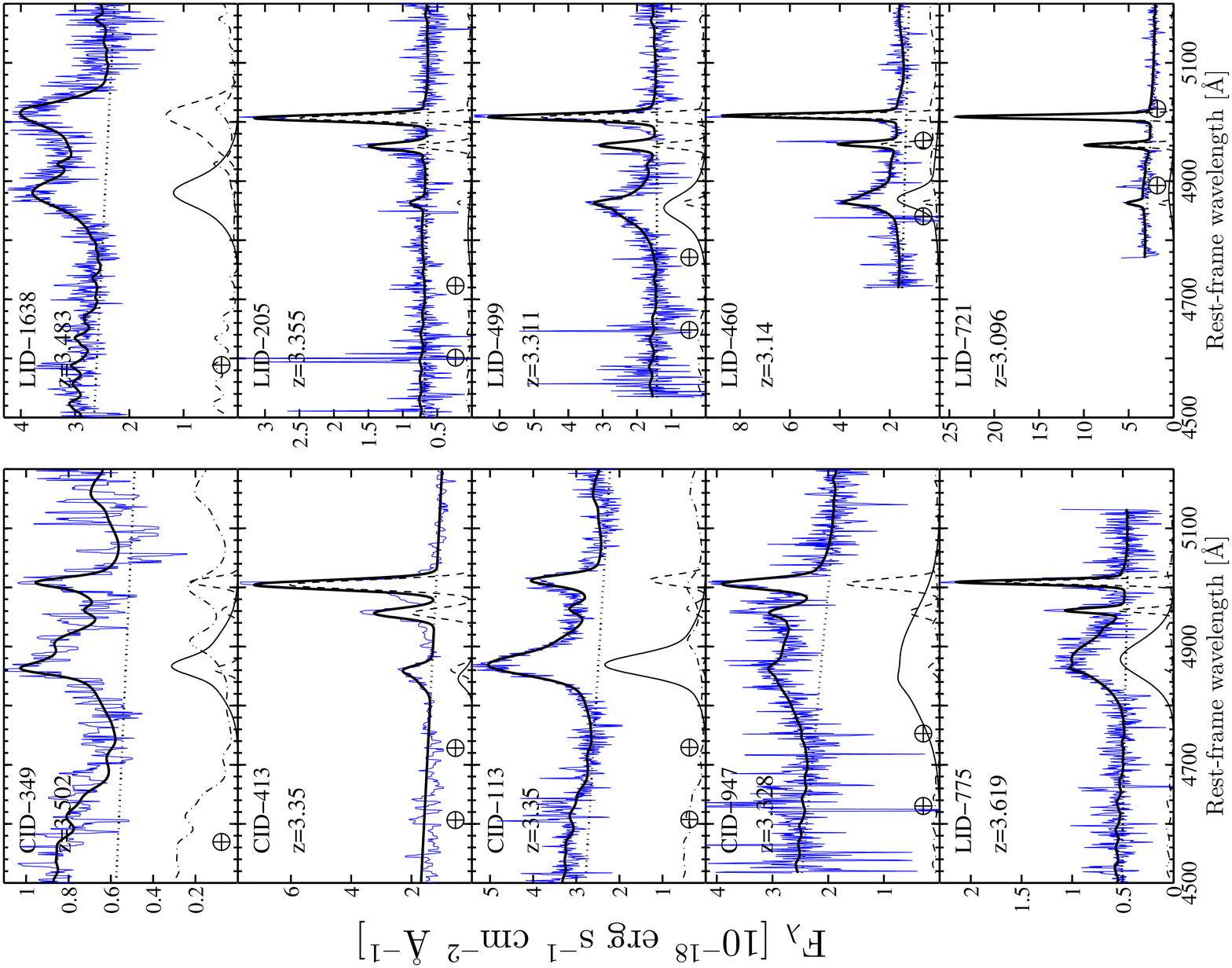} 
\caption{
Keck/MOSFIRE spectra for the \Nhb\ X-ray-selected, \ztpt\ COSMOS AGNs studied here (blue), along with the best-fitting spectral model (solid black lines).
The data are modeled with a linear continuum (dotted), a broadened iron template (dotted-dashed), and a combination of narrow (dashed) and broad (thin solid) Gaussians. 
See \autoref{subsec:analysis} for details regarding the spectral analysis.
Regions affected by telluric features are marked with encircled crosses.
The spectra are shown prior to the host-light correction. 
Note the near absence of broad \hbeta\ components in objects LID-205 and LID-721, and the peculiar broad \oiii\ profile in LID-1638 (see \autoref{subsec:em_lines}).\\
}
\label{fig:spectra_hb}
\end{figure*}

\begin{figure*}[t!]
\centering
\hspace{0.2cm}
\includegraphics[angle=-90,width=1.0\textwidth]{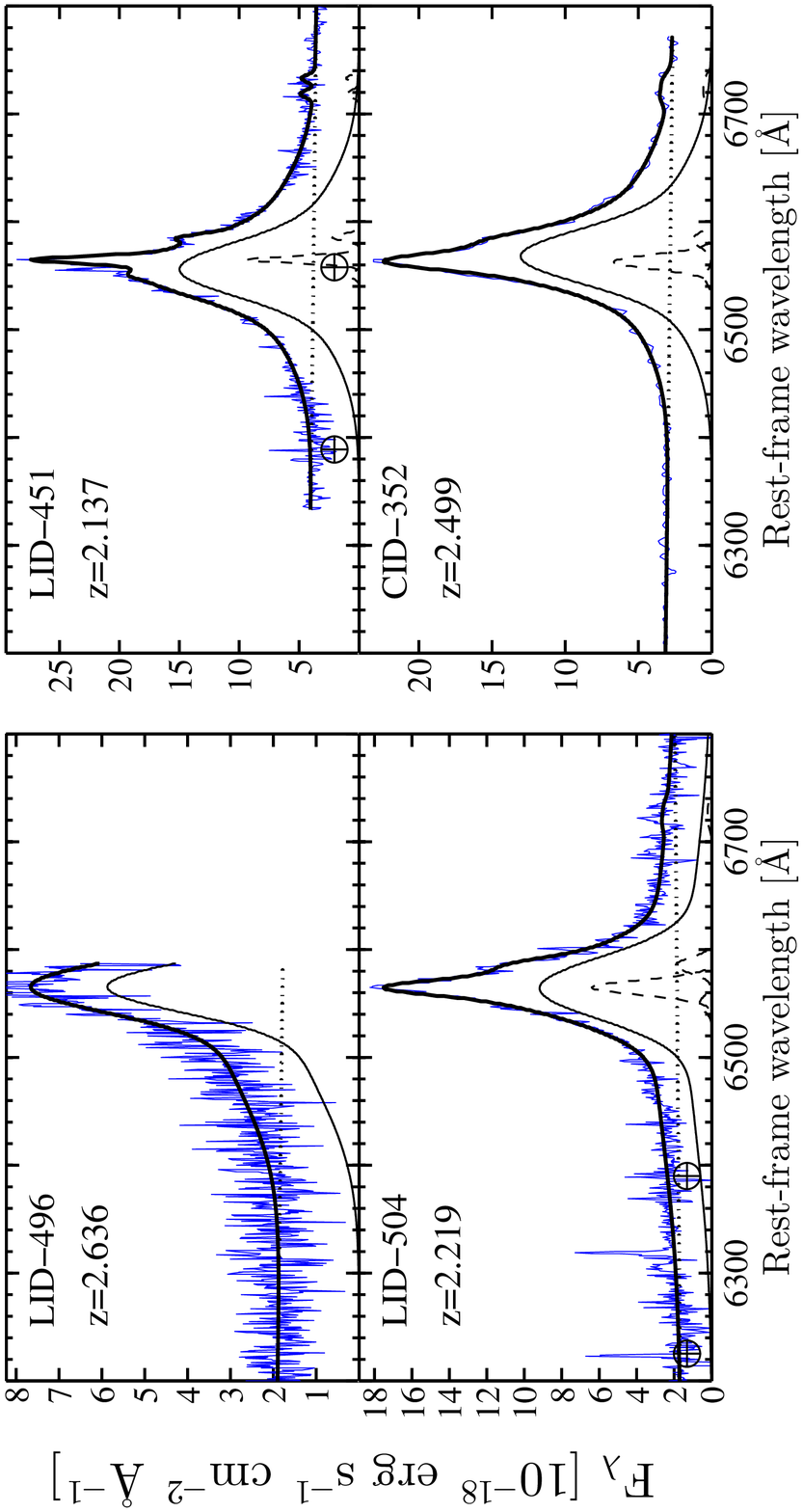} 
\caption{
Keck/MOSFIRE spectra for the \Nha\ X-ray-selected, \ztpf\ COSMOS AGNs studied here (blue), along with the best-fitting spectral model (solid black lines).
The data are modeled with a linear continuum (dotted), and a combination of narrow (dashed) and broad (thin solid) Gaussians. 
See \autoref{subsec:analysis} for details regarding the spectral analysis.
Regions affected by telluric features are marked with encircled crosses.
The spectra are shown prior to the host-light correction. 
}
\label{fig:spectra_ha}
\end{figure*}

For the source CID-349 we have combined two calibrated 1D spectra, originating from two consequent observing blocks, the second of which was considerably shorter and of poorer quality than the first one, due to varying observing conditions. 
This was done by binning the spectra in bins of 2 pixels (i.e., $\sim$1 \AA\ in the rest frame), combining the spectra through a weighted average (based on their noise spectra), and then median-smoothing the combined spectrum over 5 pixels ($\sim5$ \AA\ in the rest frame), to avoid single-pixel features inherited from the shorter and poorer-quality observing block.
Based on our experience with modeling such data, we are confident that the particular choices made in these binning and smoothing steps have little effect on the deduced spectral models and parameters, because these are mainly driven by the width of the broad Balmer lines, which is of the order of a few thousand \kms\ ($\sim$50 \AA\ in the rest frame).
Two of the fainter sources we observed (CID-955 and CID-1311) resulted in spectra that were too noisy to be used for the detailed spectral analysis required for the estimation of \mbh.
The reduced spectrum of another (optically faint) source, LID-1710, included no identifiable emission lines. 
Otherwise, the calibrated spectra of the remaining \Ntot\ sources, at both redshift bands, typically have ${\rm S/N}\sim5-7$ per instrumental spectral pixel (of about $2.2$ \AA). 
After rebinning the spectra to a uniform resolution of $1$ \AA\ in the rest frame (corresponding to $\sim45-60\,\kms$), this results in ${\rm S/N}\sim7-10$, with some of the brighter sources reaching $S/N\sim15-20$.
These (median) values of S/N per spectral bin of $1$ \AA\ (in the rest frame) are listed in \autoref{tab:obs_log}.
The final forms of the spectra of the \Ntot\ sources studied here are presented in Figures \ref{fig:spectra_hb} and \ref{fig:spectra_ha}.

\subsection{Ancillary Data}
\label{subsec:ancillary_data}

To obtain an independent constraint on intrinsic AGN-dominated luminosities, we relied on the X-ray data available for all sources from the \chandra\ catalogs in the COSMOS field (M15, \citealt{Marchesi2016_XVP_opt_ID}).
These rest-frame 2-10 \kev luminosities, \Lhard, were obtained directly from the soft-band fluxes (0.5-2 \kev), which at the redshift range of our sources probe the rest-frame hard-band (2-10 \kev) photons. 
We assumed a power-law SED with a photon index of $\Gamma=1.4$, for consistency with the analysis of the parent sample of high-redshift AGNs in the \XVP\ (M15).
As mentioned above, the X-ray luminosities we thus obtain are in the range of $\log\left(\Lhard/\ergs\right)=43.9-45$ (see \autoref{tab:lums}).
As previously noted, all the sources in our sample are robustly detected in the \emph{XMM}-COSMOS survey.
We compared the \chandra-based X-ray luminosities to those determined from the \xmm\ data, as described in \cite{Brusa2009_XCOS_hiz}.
The \chandra\ luminosities agree with the \emph{XMM} ones, with a median offset of 0.07 dex (i.e., \chandra-based luminosities are typically higher).
This difference is probably due to the different assumptions made in deriving the \emph{XMM}-based luminosities, particularly the power law of the X-ray SED ($\Gamma=1.7$ in \citealt{Brusa2009_XCOS_hiz} versus $1.4$ here).

Finally, we used data from the COSMOS/VLA radio survey \cite[][]{Schinnerer2010_VLA_COSMOS} to determine whether the sources in our sample are radio-loud (RL) AGNs.
The energy output of such RL-AGNs may be dominated by jets, and several studies have suggested that their BH masses may be systematically higher than those of the general population, perhaps due to the nature of their host galaxies \cite[e.g.,][]{McLure2004_RL}.
Four sources in our sample are robustly detected at 1.4 GHz (i.e., above $5\sigma$; CID-113, LID-1638, LID-499, and LID-451).
We calculated the radio loudness parameter, $R_{\rm L}\equiv f_\nu \left(5\,{\rm GHz}\right)/f_\nu \left({\rm optical}\right)$, following \cite{Kellermann1989}, and further assuming that the radio SED has the shape $f_\nu \propto \nu^{0.8}$.
When comparing with the rest-frame optical fluxes (either from the spectral analysis detailed in \autoref{subsec:analysis}, or the \hband\ UltraVISTA fluxes), we find that only the source LID-451 is a RL-AGN, with $R_{\rm L}\simeq117$, and the source LID-460 is marginally RL, with $R_{\rm L}\simeq10$.

\subsection{Spectral Analysis}
\label{subsec:analysis}

The spectra of the \Ntot\ sources with sufficiently high S/N were analyzed to obtain estimates of the continuum luminosity, and the luminosities and widths of the broad Balmer emission lines.
The methodology of the analysis is very similar to that discussed in numerous previous works \cite[e.g.,][and references therein]{Shang2007,Shen_dr7_cat_2011,TrakhtNetzer2012_Mg2,Mejia2015_XS_MBH} and is only briefly described here.

The spectra of the \ztpt\ sources were modeled using the procedure presented in \cite{TrakhtNetzer2012_Mg2}. 
The model consists of a linear (pseudo) continuum, a broadened iron template, and a combination of Gaussians to account for the broad and narrow emission lines, namely \heii, \hb, [O\,{\sc iii}]\,$\lambda4959$ and [O\,{\sc iii}]\,$\lambda5007$. 
The continuum flux at 5100 \AA\ was estimated directly from the best-fit linear continuum, which is performed in two narrow continuum bands, and used to measure the monochromatic continuum luminosity at (rest-frame) 5100 \AA\ (\Lop).
The broadened \feii\ template \citep{BG92} is fitted in either the 4400-4650 or 5120-5200 \AA\ spectral region,\footnote{For the two sources with $3<z<3.15$, LID-460 and LID-721, the MOSFIRE spectra do not cover the 4400-4650 \AA\ spectral region (see \autoref{fig:spectra_hb}).} and produces only negligible contamination to the 5100 \AA\ continuum band.
Most of the \ztpt\ AGNs show very low levels of \feii\ emission, although the limited quality of our spectra does not allow for a robust \emph{measurement} of the physical properties related to this emission component.
Finally, the \hb\ line is modeled with two broad Gaussian components and a single narrow one, with the latter being tied to the \oiii\ features (in terms of linewidth).
We note that the main different components are fitted in a serial manner: 
the best-fit continuum is subtracted from the original spectrum; the \feii\ template is then fitted to the continuum-free spectrum, over a different wavelength range; the best-fit \feii\ template is then subtracted, and finally the emission line model is fitted to the continuum- and iron-free spectrum.
As for the \ztpf\ sources, the \Halpha\ spectral complex was modeled using the procedure presented in \cite{Mejia2015_XS_MBH}.
The model consists of a linear (pseudo) continuum and a combination of Gaussians to account for \Halpha, [N\,{\sc ii}]\,$\lambda\lambda6548,6584$ and [S\,{\sc ii}]\,$\lambda\lambda6717,6731$. 
The \ha\ line is modeled with two broad Gaussian components and a single narrow one, again tied in width to the other nearby narrow emission lines.
The luminosity of the broad \ha\ line is calculated from the best-fit model for the broad components of the line.
All spectral fits were performed using the Levenberg--Marquardt algorithm for $\chi^2$ minimization.

For the two Balmer lines, we preferred to use \fwhm\ over \sigBLR\ as the probe of the virial velocity field of the broad-line region (BLR) gas, as the former can be more robustly estimated in spectra of moderate S/N, as is the case with our MOSFIRE data \cite[e.g.,][]{Denney2009_sys_uncer,Mejia2015_XS_MBH}.
Specifically, the study of \cite{Denney2009_sys_uncer} suggests that the use \fwhb\ may introduce biases in the estimation of \mbh\ of up to $\sim$0.1 dex, when fitting spectra with ${\rm S/N}\sim5-10$, compared to about $-0.15$ dex for \sighb. 
On the other hand, the measurement of \fwhb\ is more sensitive to the accurate removal of the narrow-line emission, with an associated mass bias of as much as an order of magnitude (in the sense of significantly underestimating \mbh), compared to $<0.2$ dex for \sighb. 
We therefore stress again that our linewidth measurements were performed for the best-fit profile of the \emph{broad} component of \hbeta, excluding the narrow-line emission, which is fitted with a separate component.
We also note that for one of the sources, LID-496, a significant fraction of the red wing of the \ha\ profile is located outside of the observed spectral range. 
To test the robustness of our fitting procedure in this case, we used a modified version of the spectrum of LID-504 that excludes the data beyond the same (rest-frame) wavelength.\footnote{We chose to use the spectrum of LID-504 since it has a similar S/N to that of LID-496, was observed within the same MOSFIRE mask, and is the next-faintest \ztpf\ source in our sample.}
The spectral parameters obtained from the simulated spectrum are in excellent agreement with those derived for the full spectrum, with differences of about 0.02, 0.02, and 0.05 dex, for \fwha, \Lha, and \Lsix, respectively.
We are therefore confident that our best-fit emission line properties are robust, within the measurement uncertainties.
The best-fit models are shown in Figures \ref{fig:spectra_hb} and \ref{fig:spectra_ha}.

We derived measurement-related uncertainties on the best-fit Balmer line properties using a resampling approach.
For each of the spectra, we generated a series of 100 realizations of the data, each of which differed from the observed spectral data by a random, normally distributed offset, determined from the error spectrum of that source. 
Each of these realizations was modeled using the aforementioned line fitting procedures, and the relevant quantities were recorded.
Thanks to the high-quality MOSFIRE data, we obtain relatively small measurement-related uncertainties on the quantities of interest (luminosities and linewidths).
The typical uncertainty on \Lop\ (among the \ztpt\ sources) is below 0.05 dex, which is smaller than the uncertainty imposed by the flux calibration.
The typical uncertainty of the broad-line \fwhm\ is a few hundred \kms.
When combining these quantities to derive ``virial'' mass estimators, the resulting uncertainties are of order 0.1 dex, which is smaller than the systematic uncertainties (see details in \autoref{subsec:est_mbh_lledd_tg}).
%

\capstartfalse
\begin{deluxetable*}{llccccccccc}
\tabletypesize{\footnotesize}
\tablecolumns{10}
\tablewidth{0pt}
\tablecaption{Redshifts and Multi-wavelength Luminosities \label{tab:lums}}
\tablehead{
  \colhead{Subsample} &
  \colhead{Object ID}  &
  \colhead{$z$}  &
  \colhead{$z_{\rm NIR}$\tablenotemark{a}}  &
  \colhead{$\log$ \Luv \tablenotemark{b}} &
  \colhead{$M_{1450}$ \tablenotemark{c}} &
  \colhead{$\log$ \Lhard \tablenotemark{d}} & 
  \colhead{$\log$ \Lop \tablenotemark{e}} & 
  \colhead{$f_{\rm AGN,5100}$ \tablenotemark{f}} & 
  \multicolumn{2}{c}{$\log$ \Lbol\ (\ergs) \tablenotemark{g}} \\
   & &  &  & (\ergs) & &  (\ergs) & (\ergs) & &  opt. & X-ray
}
\startdata
\ztpt\ & CID-349 & $ 3.5150 $ & $ 3.5017 $ & $ 45.43 $ & $ -23.69 $ & $ 44.44 \pm 0.07 $ & $ 45.11 ^{+0.006}_{-0.008}$ & $ 0.72 $ & $ 45.79 $ & $ 46.11	$ \\
~~~    & CID-413 & $ 3.3450 $ & $ 3.3504 $ & $ 45.06 $ & $ -22.77 $ & $ 44.53 \pm 0.06 $ & $ 45.38 ^{+0.008}_{-0.009}$ & $ 0.58 $ & $ 45.96 $ & $ 46.23	$ \\
~~~    & CID-113 & $ 3.3330 $ & $ 3.3496 $ & $ 46.08 $ & $ -25.33 $ & $ 44.64 \pm 0.05 $ & $ 45.71 ^{+0.004}_{-0.004}$ & $ 1.00 $ & $ 46.49 $ & $ 46.37	$ \\
~~~    & CID-947 & $ 3.3280 $ & $ 3.3279 $ & $ 45.91 $ & $ -24.90 $ & $ 43.86 \pm 0.16 $ & $ 45.55 ^{+0.009}_{-0.009}$ & $ 0.86 $ & $ 46.34 $ & $ 45.35	$ \\
~~~    & LID-775 & $ 3.6260 $ & $ 3.6193 $ & $ 45.64 $ & $ -24.23 $ & $ 44.65 \pm 0.06 $ & $ 45.10 ^{+0.037}_{-0.047}$ & $ 0.67 $ & $ 45.75 $ & $ 46.40	$ \\
~~~    & LID-1638& $ 3.5030 $ & $ 3.4827 $ & $ 45.75 $ & $ -24.49 $ & $ 44.47 \pm 0.07 $ & $ 45.77 ^{+0.013}_{-0.008}$ & $ 0.77 $ & $ 46.44 $ & $ 46.15	$ \\
~~~    & LID-205 & $ 3.3560 $ & $ 3.3552 $ & $ 45.62 $ & $ -24.17 $ & $ 44.75 \pm 0.04 $ & $ 45.15 ^{+0.012}_{-0.028}$ & $ 0.65 $ & $ 45.78 $ & $ 46.53	$ \\
~~~    & LID-499 & $ 3.3020 $ & $ 3.3114 $ & $ 44.91 $ & $ -22.41 $ & $ 44.47 \pm 0.08 $ & $ 45.49 ^{+0.017}_{-0.022}$ & $ 0.71 $ & $ 46.14 $ & $ 46.15	$ \\
~~~    & LID-460 & $ 3.1430 $ & $ 3.1401 $ & $ 44.90 $ & $ -22.38 $ & $ 44.99 \pm 0.03 $ & $ 45.37 ^{+0.003}_{-0.007}$ & $ 0.64 $ & $ 45.98 $ & $ 46.84	$ \\
~~~    & LID-721 & $ 3.1080 $ & $ 3.0959 $ & $ 46.11 $ & $ -25.40 $ & $ 44.53 \pm 0.04 $ & $ 45.58 ^{+0.010}_{-0.005}$ & $ 0.66 $ & $ 46.20 $ & $ 46.23	$
\\\hline \\ 
    &  &   &   &     &     &    & $\log\,L_{6200}$ & $f_{\rm AGN,6200}$ & (\Lha) &   \\
\hline \\ [-1.75ex]
\ztpf\ & LID-496 & $ 2.6300 $ & $ 2.6360 $ & $ 45.84 $ & $ -24.71 $ & $ 44.29 \pm 0.08 $ & $ 44.82 ^{+0.006}_{-0.004}$ & $ 1.00 $ & $ 45.66 $ & $ 45.91	$ \\
~~~    & LID-504 & $ 2.2220 $ & $ 2.2191 $ & $ 45.24 $ & $ -23.22 $ & $ 44.95 \pm 0.05 $ & $ 44.59 ^{+0.030}_{-0.029}$ & $ 0.73 $ & $ 45.71 $ & $ 46.79	$ \\
~~~    & LID-451 & $ 2.1220 $ & $ 2.1367 $ & $ 45.67 $ & $ -24.30 $ & $ 44.61 \pm 0.04 $ & $ 44.93 ^{+0.004}_{-0.002}$ & $ 0.97 $ & $ 45.81 $ & $ 46.33	$ \\
~~~    & CID-352 & $ 2.4978 $ & $ 2.4993 $ & $ 46.13 $ & $ -25.44 $ & $ 44.88 \pm 0.03 $ & $ 44.99 ^{+0.004}_{-0.003}$ & $ 1.00 $ & $ 45.89 $ & $ 46.70	$ 
\enddata
\tablenotetext{a}{Redshift measured from the best-fit model of the \oiii\ or (narrow) \Halpha\ lines.}
\tablenotetext{b}{Monochromatic luminosity at rest-wavelength 1450\AA, obtained from the optical spectra (see \autoref{tab:obs_log}).}
\tablenotetext{c}{Absolute magnitude at 1450\AA, following $M_{1450}=-2.5\log\Luv+89.9$.}
\tablenotetext{d}{\chandra-based, obscuration-corrected rest-frame hard-band [$\left(2-10\right)\,\kev$] luminosity, taken from \cite{Marchesi2015_XVP_hiz}.}
\tablenotetext{e}{Monochromatic luminosities at rest-wavelength 5100 \AA\ (for \ztpt\ AGNs) or 6200\AA\ (for \ztpf\ AGNs), \emph{uncorrected} for host contamination. The tabulated errors reflect only measurement-related uncertainties.} 
\tablenotetext{f}{AGN luminosity fraction at $\lambda_{\rm rest}=5100$ or $6200$\AA, determined from SED decomposition.} 
\tablenotetext{g}{Bolometric luminosity estimates based either on \Lop\ (or \Lha) or on \Lhard.}
\end{deluxetable*}
\capstarttrue

\subsection{Derivation of \Lbol, \mbh\ and \lledd}
\label{subsec:est_mbh_lledd_tg}

The bolometric luminosities of the sources, \Lbol, were estimated in several different ways.
First, for consistency with previous studies of high-redshift unobscured AGNs with \mbh\ estimates, we applied bolometric corrections that translate the optical continuum and \ha\ line luminosities to bolometric luminosities (i.e., \fbol).
For \fbolopt, we used the luminosity-dependent prescription described in \cite{TrakhtNetzer2012_Mg2}, which in turn relies on the $B$-band bolometric corrections presented in \cite{Marconi2004}, and translated to 5100 \AA\ assuming a UV--optical SED with $f_{\nu}\propto \nu^{-1/2}$ \cite[][]{VandenBerk2001}.
In the relevant range of \Lop, these corrections can be described by 
\begin{equation}
\fbolopt =  6.58 -0.89 {\cal L}_{5100,45} + 0.22 {\cal L}_{5100,45}^2   \, , 
\label{eq:fbol_opt_poly}
 \end{equation}
where ${\cal L}_{5100,45}\equiv \log\left(\Lop/10^{45}\,\ergs\right)$.
For the \ztpf\ objects, we used the \Lha-dependent bolometric corrections suggested in \cite{Greene2007_BHMF}, which provide
\begin{equation}
\Lbol\left(\Lha\right) =  2.34\times10^{44}\, \left(\frac{\Lha}{10^{42}\,\ergs}\right)^{0.86}   \, . 
\label{eq:Lbol_Lha}
 \end{equation}
This \Lha-based prescription was calibrated against \Lop\ for a sample of low-redshift AGNs, assuming $\fbolopt=9.8$. 
To test the consistency of these \Lha-based estimates of bolometric luminosity, we translated the observed continuum luminosities at 6200 \AA\ (\autoref{tab:lums}) to \Lop\ (assuming $f_{\nu}\propto \nu^{-1/2}$) and then used \autoref{eq:fbol_opt_poly} to obtain another set of \Lbol\ estimates for the  \Nha\ \ztpf\ sources. 
These latter \Lsix-based estimates of \Lbol\ are consistent with those derived directly from \autoref{eq:Lbol_Lha}, with a (median) offset of merely 0.03 dex.
The bolometric luminosities obtained through Equations \ref{eq:fbol_opt_poly} and \ref{eq:Lbol_Lha} are in the range of  $\Lbol\simeq\left(6-31\right)\times10^{45}\,\ergs$.
Second, we used the X-ray luminosities measured from the \chandra\ data, and X-ray bolometric corrections. 
For $\fbol\left(\Lhard\right)$, we used the prescription of \cite{Marconi2004}, for consistency with other studies using the \chandra\ survey data. 
These \chandra-based \Lbol\ values are in the range $\Lbol\left(\Lhard,Chandra\right)=\left(2-68\right)\times10^{45}\,\ergs$.
Since the \emph{XMM}-based estimates of \Lhard\ are highly consistent with the \chandra\ ones, they result in similar X-ray-based \Lbol\ estimates.
Finally, we note that yet another set of \Lbol\ estimates for nine of our sources (six of those at \ztpt) is available from the multi-wavelength analysis performed for our sources as part of the \emph{XMM}-COSMOS survey by \citet[][see also \citealt{Lusso2011_Lbol}]{Lusso2010_XCOS_type1}.
Unlike the previous \Lbol\ estimates discussed here, these were obtained by integrating the multi-wavelength AGN SEDs up to 1 \mic\ (and further fixing the unobserved FUV and hard X-ray parts of the SED).
These \emph{XMM}- and SED-based \Lbol\ estimates are in the range $\Lbol\left({\rm SED},XMM\right)=\left(1-41\right)\times10^{45}\,\ergs$, and in good agreement with our estimates of \Lbol\ based on \Lop\ or \Lha\ -- the median offset is 0.09 dex (0.03 dex for the \ztpt\ AGNs; \emph{XMM}- and SED-based \Lbol\ estimates are higher), and the scatter is 0.37 dex (0.17 dex for \ztpt\ AGNs).
In \autoref{tab:lums} we list the different bolometric luminosities we obtained for our sources.
The \Lhard-based estimates of \Lbol\ for our sources are generally consistent with those derived from \Lop\ and \Lha, with a median offset of about 0.07 dex between the latter and the former, and virtually all the sources having differences within 0.5 dex. 
The extreme source \mysobj\ is exceptionally weak in the X-rays, resulting in an \Lbol\ difference of almost an order of magnitude. 
Moreover, as noted in T15, the X-ray luminosity of this broad-absorption-line quasar as derived from the \emph{XMM}-COSMOS survey is significantly higher than that obtained from the \chandra\ data, which might be related to varying obscuration along the line of sight. 
In what follows, we chose to use the  bolometric luminosities based on \Lop\ and \Lha, given the (generally) higher quality of the rest-frame optical data, the limited availability of other \Lbol\ estimates (i.e., \emph{XMM}+SED), and in order to be consistent with previous studies of $z>2$ unobscured AGNs (see the comparison samples in \autoref{subsec:dist_mbh_lledd}).

We estimated black hole masses for the sources using the quantities derived from the best-fitting spectral models, and following the prescription used in several recent works \cite[][]{Netzer2007_MBH,TrakhtNetzer2012_Mg2}.
For the \ztpt\ sources, we correct the continuum luminosities to account for the emission from the stellar component in the host galaxies. 
These scaling corrections are derived from the spectral compositions of the broad-band SEDs of the sources, which are described in detail in a forthcoming publication. 
In short, the stellar component is modeled using a large grid of (single) stellar population models, with a broad range of ages, star formation histories, and dust extinction. 
We use the stellar template that provides the best fit to the SED, provided that the UV-optical regime of all SEDs is AGN-dominated.
The scaling factors thus computed, which are simply the fraction of AGN-related emission at $\lambda_{\rm rest}=5100$ \AA, are in the range of $f_{\rm AGN}\left(5100{\rm \AA}\right)\sim0.55-1$.
Next, \hb-based BH masses are estimated using the expression
\begin{eqnarray}
  \mbh\left(\hb\right)=1.05\times 10^8  
			\left(\frac{L_{5100}}{10^{46}\,\ergs}\right)^{0.65} \nonumber \\
			\times \left[\frac{\fwhb}{10^3\,\kms}\right]^2 \,\, \Msun \,\, .
\label{eq:M_Hb}
\end{eqnarray}
This prescription is based on the $\RBLR-\Lop$ relation obtained through reverberation mapping of low-redshift sources with comparable (optical) luminosities \cite[][]{Kaspi2005}, and assumes a BLR ``virial factor'' of $f=1$ \cite[see also][]{Onken2004,Woo2010_LAMP_Msig,Grier2013_sigs_PGs}.
The exponent of the luminosity term means that the aforementioned host-light corrections affect the derived masses by at most $\sim$0.17 dex.
We verified that using alternative \RBLR\ estimators would not significantly affect our determinations of \mbh.
In particular, in the range of optical luminosities of our sources, the $\RBLR-\Lop$ relation of \cite{Bentz2013_lowL_RL} results in BLR sizes (and therefore BH masses) that are systematically \emph{smaller} than those derived by the relation of \cite{Kaspi2005}. 
The difference between the two \RBLR\ estimates increases with increasing \Lop\ (or \mbh), but for our sources it remains very small, in the range 0.02-0.1 dex (median value 0.06 dex).  

For the sources at \ztpf\ we estimated \mbh\ from the luminosity and width of the \Halpha\ line, following the prescription of \cite{Greene_Ho_Ha_2005}:
\begin{eqnarray}
  \mbh\left(\ha\right)=1.3\times 10^6  
			\left(\frac{L_{\ha}}{10^{42}\,\ergs}\right)^{0.57} \nonumber \\
			\times \left[\frac{\fwha}{10^3\,\kms}\right]^{2.06} \,\, \Msun \,\, .
\label{eq:M_Ha}
\end{eqnarray}
This \mbh\ was derived through an empirical secondary calibration against \hbeta-related quantities (\Lop\ and ${\rm FWHM}\left[\hb\right]$).\footnote{Thus, the luminosity-term exponent (0.57) is not directly observed in an $\RBLR-\Lha$ relation, and the velocity-term exponent (2.06) is not strictly virial.}
These two prescriptions were also used to derive masses for each of the spectra simulated within our resampling scheme, thus providing measurement-related uncertainties on the \mbh\ estimates.

We note that the relevant luminosities of our sources are well within the range of the reverberation mapping campaigns that stand at the base of ``virial'' estimates of \mbh. In particular, our \ztpt\ sources have (host-corrected) optical luminosities comparable with those of low-redshift PG quasars, for which \RBLR\ estimates were obtained in several reverberation mapping studies \cite[e.g.,][]{Kaspi2000,Kaspi2005,Vester_Peterson2006}.
Thus, our virial estimates of \mbh\ do not require the extrapolation of the $\Lop-\RBLR$ relation toward extremely high luminosities, which is often the case in other studies of $z\gtrsim2$ AGNs 
\cite[e.g.,][]{Shemmer2004, Marziani2009}.

The \mbh\ and \Lbol\ estimates were finally combined to obtain Eddington ratios, $\lledd \equiv \Lbol / \left(1.5\times10^{38} \, \mbh/\Msun\right)$ (suitable for solar-metalicity gas).
As mentioned above, we choose to use the \Lop-based estimates of \Lbol. Choosing instead the \Lhard-based estimates would lead to slightly higher values of \lledd. 
Such a choice would not significantly affect any of our main findings, and would actually strengthen our claim of a lack of low-\lledd\ and high-\mbh\ AGNs (see \autoref{subsec:dist_mbh_lledd}).
Our estimates of \mbh\ and \lledd\ are listed in \autoref{tab:fit_pars}.
Since the measurement-related uncertainties on \mbh\ are relatively small, rarely exceeding 0.1 dex, the real uncertainties on \mbh\ are dominated by the systematics associated with the ``virial'' mass estimators we used. 
These are estimated to be of order $\sim$0.3 dex for the \ztpt\ sources \cite[e.g.,][]{Shen2013_rev}, and yet higher for the \ztpf\ ones, as their mass estimator is based on a secondary calibration of $\mbh\left(\Halpha\right)$.

\capstartfalse
\begin{deluxetable*}{llccccccccc}
\tablecolumns{11}
\tablewidth{0pt}
\tablecaption{Spectral Measurements and Derived SMBH Properties \label{tab:fit_pars}}
\tablehead{
  \colhead{sub-sample} &
  \colhead{Object ID}  &
  \colhead{$\log$ \Lhb} &
  \colhead{\fwhb} &
  \colhead{$\log$ \mbh} &
  \colhead{$\log$ \lledd \tablenotemark{a}} &
  \multicolumn{2}{c}{\Maddot\ (\mpyr) \tablenotemark{b} } &
  \multicolumn{2}{c}{\tgrow\ (Gyr) \tablenotemark{c}} \\
   & & (\ergs) &  (\kms) &  (\Msun) &    & \Lbol\ & AD & \lledd\ & $\dot{M}$ }
\startdata
\ztpt\ & CID-349 & $ 43.14 $ & $  3223^{+ 592}_{ -385} $ & $ 8.37^{+0.13}_{-0.11} $ & $ -0.76 $ & $  1.08 $ & $  1.29 $ & $  0.25 $ & $  0.20 $ \\
~~~    & CID-413 & $ 42.85 $ & $  4149^{+1707}_{-1143} $ & $ 8.70^{+0.18}_{-0.25} $ & $ -0.92 $ & $  1.60 $ & $  1.11 $ & $  0.37 $ & $  0.51 $ \\
~~~    & CID-113 & $ 43.80 $ & $  2959^{+ 101}_{ -117} $ & $ 8.78^{+0.03}_{-0.03} $ & $ -0.46 $ & $  5.51 $ & $  6.54 $ & $  0.13 $ & $  0.10 $ \\
~~~    & CID-947 & $ 43.52 $ & $ 11330^{+ 929}_{ -799} $ & $ 9.84^{+0.07}_{-0.06} $ & $ -1.67 $ & $  3.03 $ & $  0.22 $ & $  2.09 $ & $  34.68$ \\
~~~    & LID-775 & $ 43.40 $ & $  4700^{+ 450}_{ -328} $ & $ 8.67^{+0.10}_{-0.06} $ & $ -1.10 $ & $  0.99 $ & $  0.56 $ & $  0.55 $ & $  0.92 $ \\
~~~    & LID-1638& $ 43.67 $ & $  4071^{+ 316}_{ -308} $ & $ 9.02^{+0.06}_{-0.07} $ & $ -0.75 $ & $  4.86 $ & $  3.09 $ & $  0.25 $ & $  0.37 $ \\
~~~    & LID-499 & $ 43.54 $ & $  3451^{+ 606}_{ -360} $ & $ 8.67^{+0.15}_{-0.10} $ & $ -0.70 $ & $  2.43 $ & $  2.32 $ & $  0.23 $ & $  0.22 $ \\
~~~    & LID-460 & $ 43.52 $ & $  2260^{+  45}_{  -89} $ & $ 8.19^{+0.02}_{-0.05} $ & $ -0.39 $ & $  1.70 $ & $  3.94 $ & $  0.11 $ & $  0.04 $ \\
\hline\\
    &  &  $\log$ \Lha\ & \fwha\ &     &       &            &           &          & \\
\hline \\ [-1.75ex]
\ztpf\ & LID-496 & $ 43.50 $ & $  3533^{+  53}_{  -39} $ & $ 8.10^{+0.02}_{-0.01} $ & $ -0.61 $ & $  0.81 $ & $  2.15 $ & $  0.18 $ & $  0.07 $ \\
~~~    & LID-504 & $ 43.56 $ & $  3401^{+ 148}_{ -100} $ & $ 8.10^{+0.06}_{-0.05} $ & $ -0.56 $ & $  0.91 $ & $  0.59 $ & $  0.16 $ & $  0.24 $ \\
~~~    & LID-451 & $ 43.67 $ & $  3278^{+  71}_{ -139} $ & $ 8.13^{+0.01}_{-0.06} $ & $ -0.50 $ & $  1.14 $ & $  2.75 $ & $  0.14 $ & $  0.05 $ \\
~~~    & CID-352 & $ 43.77 $ & $  3261^{+ 236}_{ -279} $ & $ 8.18^{+0.06}_{-0.07} $ & $ -0.46 $ & $  1.38 $ & $  3.13 $ & $  0.13 $ & $  0.05 $ 
\enddata
\tablenotetext{a}{Based on \Lbol\ estimated from \Lop\ (or \Lha).}
\tablenotetext{b}{Accretion rate estimates based on either \Lbol\ (and $\eta=0.1$), or an accretion disk model \autoref{eq:mdot_ad} (``AD'').}
\tablenotetext{c}{Based on either \lledd\ (via \autoref{eq:tau_salpeter}) or on \Maddot, and further assumes $\eta=0.1$.}
\end{deluxetable*}
\capstarttrue

\vspace{0.2cm}
\section{Results and Discussion}
\label{sec:res_and_disc}

We next discuss the main results of the detailed analysis of the Balmer emission line complexes. 
We first highlight a few objects with peculiar emission line properties, before addressing the implications of our measurements for the observed early evolution of SMBHs.

\subsection{Emission Line Properties}
\label{subsec:em_lines}

Two of the \ztpt\ sources, LID-205 and LID-721, have extremely weak or indeed undetectable broad \hb\ emission lines.
Our fitting procedure suggests that the rest-frame equivalent widths of these components are approximately $\ewhb\simeq10-15$ \AA. 
More importantly, a series of (manual) fitting attempts demonstrated that the data can be adequately modeled without \emph{any} broad \hb\ components.
We also verified that these low \ewhb\ values are not due to measurement-related uncertainties.
For LID-205, 90\%\ (99\%) of the resampling simulations resulted in $\ewhb<18$ \AA\ (30 \AA, respectively).
For LID-721, the corresponding quantiles are $\ewhb<20$ and $25$ \AA, respectively.
The best-fit values are lower, by at least a factor of 4, than the median value of \ewhb\ we find for the rest of the \ztpt\ sources. 
Moreover, such weak \hb\ lines are not observed at all within other samples of $z\gtrsim2$ AGNs \cite[][]{Shemmer2004,Netzer2007_MBH,Marziani2009}, where the weakest lines have $\ewhb\sim40$\AA, and the median values are above $\sim75$\AA.
Another \ztpt\ source, CID-413, has a relatively weak broad \hb\ line, with $\ewhb=31$ \AA. 
Our simulations, however, show that the \hb\ emission can be accounted for with significantly stronger components, reaching $\ewhb\simeq70$ \AA. 
Indeed, this ambiguity regarding the broad component of CID-413 is reflected in the atypically large uncertainties on \fwhb\ and \mbh\ (see \autoref{tab:fit_pars}).
We chose, however, to include this source in the analysis that follows, since even the most extreme realizations present $\ewhb>25$ \AA.

We stress that the two ``\hb-weak'' sources we identified have strong and unambiguous \oiii\ emission lines, with flux ratios $\oiii/\hb
\gg3$, further supporting the identification of the sources as emission line systems dominated by an AGN ionization field \cite[e.g.,][]{BPT1981,Kewley2006}.
We verified that the observed-frame optical, rest-frame UV zCOSMOS and IMACS spectra of the two \hb-weak AGNs present broad and strong high-ionization \CIV\ emission lines.
Indeed, the \civ\ lines have ${\rm EW}\left(\civ\right)=118$ and $57$ \AA\ (for LID-205 and LID-721, respectively). 
This, as well as the strong \oiii\ lines, suggests that the low EWs of \hbeta\ are \emph{not} due to attenuation by dust along the line of sight.
Furthermore, the ratio of UV to optical luminosities of the \hb\ weak AGNs, $\Luv/\Lop\simeq3$, is consistent with what is found in large samples of normal AGNs \cite[e.g.,][]{TrakhtNetzer2012_Mg2}, suggesting that the broad \hb\ lines in these two sources are emph{not} diluted by stellar emission from the host.
We also note that the broad \hb\ lines in these sources are significantly weaker than those detected in the spectra of ``weak line quasars'', which are defined based on their weak UV lines (i.e., $\Lya+\nv$, or \civ; see, e.g., \citealt{Shemmer2010_WLQs,Plotkin2015_WLQs}, and references therein).
One intriguing explanation may be that the \hb-weak AGNs have experienced a dramatic decrease in the emission of ionizing radiation since the optical spectra were taken, i.e. on a roughly year-long timescale (in the AGN reference frames). 
This change may have driven a sharp decrease in the BLR emission, but has yet to reach the more extended NLR, which would explain the strong \oiii\ emission.
Such a drastic decrease in ionizing flux should, however, manifest itself also as a decrease in (rest-frame) optical continuum luminosity, 
which is not observed (see the comparison of \kband\ fluxes in \autoref{tab:obs_log}).
In this sense, our \hb-weak AGNs are inconsistent with the growing number of ``changing-look'' AGNs, detected through dramatic drops in both UV-optical continuum \emph{and} BLR emission \cite[see, e.g., recent studies by][and references therein]{Denney2014_Mrk590,LaMassa2015_changing,Runnoe2016_changing}.
In any case, revisiting these sources with optical spectroscopy may test this explanation and clarify the situation.
We therefore conclude that our sample contains two sources (about 12.5\% of the sample) with abnormally weak broad \hb\ lines, which are not due to the lack of gas in the BLR.

The spectrum of one other \ztpt\ source, LID-1638, presents an abnormally broad \oiii\ emission feature. A manual inspection of the data provides a rough estimate of ${\rm FWHM}\sim3000\,\kms$ for the width of this feature. 
At these large widths, the feature is basically a combination of the two different \oiii\ emission lines (with some additional, minor contribution from \feii). 
This width appears to be comparable to that of the adjacent \hb\ line, which otherwise appears rather normal.
Such broad \oiii\ emission features are rarely reported in large samples of lower-redshift AGNs \cite[e.g.,][]{BG92,Shen_dr7_cat_2011,TrakhtNetzer2012_Mg2},
but may be related to prominent blue wings \cite[e.g.,][]{Komossa2008}.\footnote{The automated procedures used for very large surveys (e.g., SDSS) are restricted to $\fwhm\simeq1000\,\kms$ and obviously lack a manual inspection of the (tens of thousands of) spectra.} 
Another explanation is that the \oiii\ profile consists of two separate narrow lines, emitted from separate NLRs, as observed in dual AGN candidates \cite[e.g.,][and references therein]{Comerford2012}. 
In any case, a detailed analysis and interpretation of the peculiar \oiii\ profile are beyond the scope of the present study, as we focus on the broad \hbeta\ component.
To account for the broadened \oiii\ emission, we refitted the spectrum of this source with a modified constraint of ${\rm FWHM}\leq3000\,\kms$ for the narrow emission features (both \oiii\ and \hb). 
The \fwhb\ resulting from this, of about 4100 \kms, is highly consistent with the value obtained with the ``standard'' line fitting procedure. 
Removing the width constraint altogether results in yet broader \oiii\ features, exceeding 5000 \kms, but with \fwhb\ decreasing to $\sim3700\,\kms$. 
This is mainly due to the fact that the fitting procedure does not allow for a significant (broader than usual) narrow component for \hb.
However, we find the overall fit to the data in this case unsatisfactory, and note that in any case this would result in a decrease of merely 0.1 dex in \mbh. 
The best-fit parameters tabulated for LID-1638 in \autoref{tab:fit_pars} are therefore those obtained with the constraint ${\rm FWHM} \oiii \leq 3000\,\kms$.

\begin{figure}
\centering
\includegraphics[angle=-90, width=0.48\textwidth]{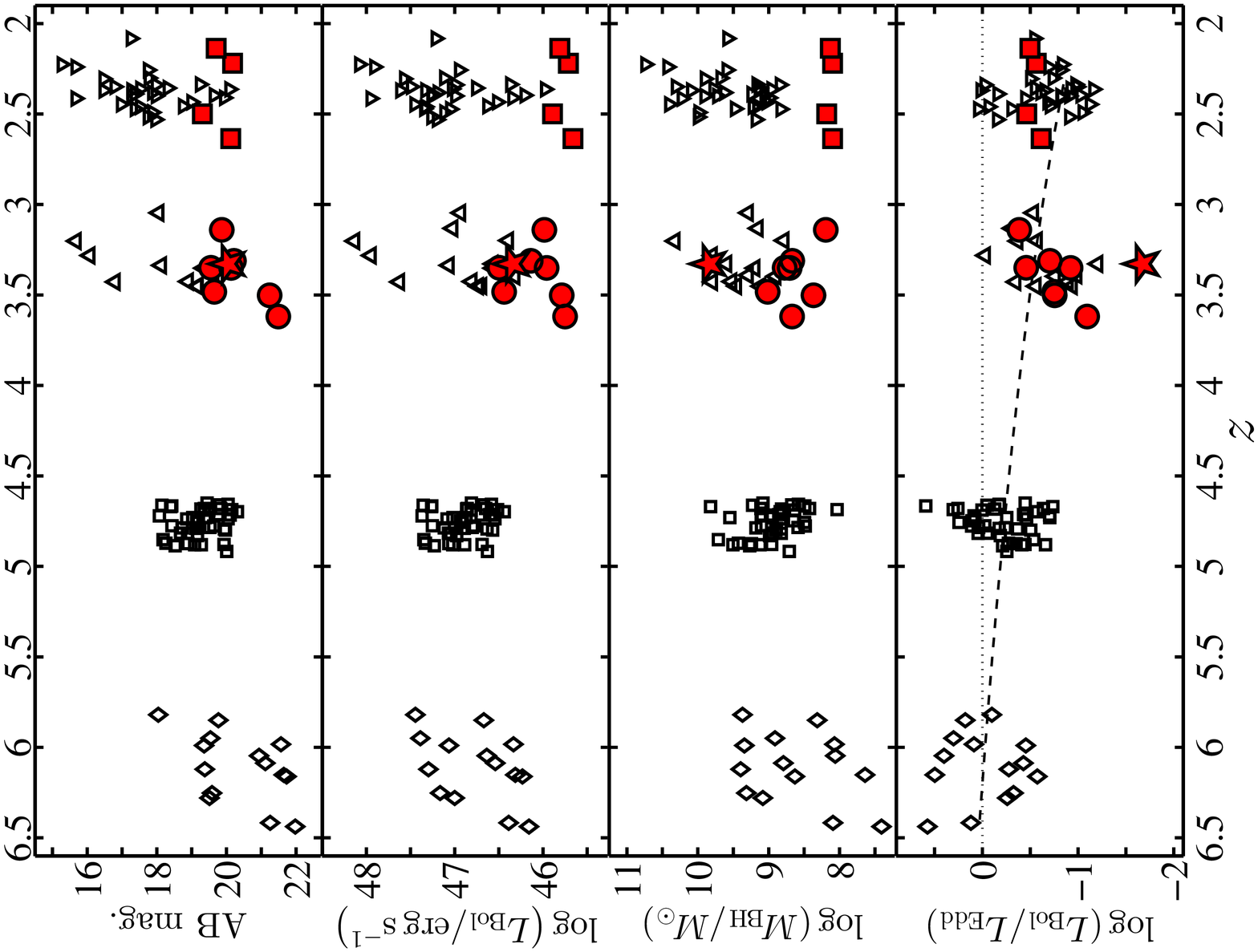} 
\caption{
From top to bottom, trends of observed (NIR) brightness, \Lbol, \mbh, and \lledd\ for the available samples of unobscured AGNs at $z>2$, with reliable determinations of \mbh.
The red symbols represent the measurements reported in this work, at \ztpt\ and $2.4$ (circles and squares, respectively).
\mysobj, which was analyzed in detail in \cite{Trakhtenbrot2015_CID947}, is highlighted by a star.
The different black symbols represent other, optically selected sources, studied in the combined sample of \citet[][]{Shemmer2004} and \citet[][triangles at \znetprev]{Netzer2007_MBH}; \citet[][squares at \zfpe]{Trakhtenbrot2011}; and the combined samples of \citet[][]{Kurk2007} and \citet[][diamonds at \zsix]{Willott2010_MBH}.
The dotted line in the bottom panel marks the Eddington limit, i.e., $\lledd=1$.
The dashed line follows $\lledd\propto\left(1+z\right)^2$, reaching $\lledd=1$ at $z=6.2$, which represents the general trend among the samples considered here.
}
\label{fig:L_M_Ledd_vs_z}
\end{figure} 

\subsection{Trends in \mbh\ and \lledd\ at $z>2$}
\label{subsec:dist_mbh_lledd}

\autoref{fig:L_M_Ledd_vs_z} presents the distributions of relevant apparent brightness and estimates of \Lbol, \mbh, and \lledd\ for the sources studied here, as a function of redshift, in the context of other samples of optically selected and unobscured AGNs at $z>2$, for which these quantities were reliably determined. 
The relevant samples are those presented by \citet{Shemmer2004} and \citet[at \ztpt\ and $2.4$]{Netzer2007_MBH}; by \citet[\zfpe]{Trakhtenbrot2011}; and by \cite{Kurk2007} and \citet[\zsix]{Willott2010_MBH}.
The apparent magnitudes in the top panel of the diagram represent the NIR bands at which either the \hb\ (\znetprev) or \mgii\ broad emission lines would be observed, which is the $H$-band for $z\simeq2.4$ and $4.8$ sources or the $K$-band for \ztpf\ and $6.2$ sources.\footnote{Note that for our \ztpf\ COSMOS AGNs we use the \hband\ magnitudes \cite[from UltraVISTA][]{McCracken2012_COSMOS_UltraVISTA}, although we study the \halpha\ line in the $K$ band.
The magnitudes for the other sources were compiled from the original studies, where the $K$-band magnitudes of the \zsix\ sources were estimated from the published $J$-band magnitudes \cite[][]{Jiang2006_Spitzer}, and assuming $J_{\rm Vega}-K_{\rm Vega}=1.25$ and $H_{\rm Vega}-K_{\rm Vega}=0.75$.
}
The \hbeta-based \mbh\ estimates for all the \znetprev\ AGNs in these comparison samples are based on the same prescription as we use here (\autoref{eq:M_Hb}).
For consistency with previous studies (and in particular with \citealt{Trakhtenbrot2011}), the \mgii-based \mbh\ estimates for $z>4.5$ sources are based on the calibration of \cite{McLure_Dunlop2004}.
The bolometric corrections for all the comparison sources are based on the same procedure as the one used here (\autoref{eq:fbol_opt_poly}), extended to \fbolthree\ for $z>4.5$ sources \cite[see][]{TrakhtNetzer2012_Mg2}.
We note that several other studies have provided (relatively small) samples with \mbh\ estimates for AGNs at $2 \lesssim z\lesssim3$ \cite[e.g.,][]{Alexander2008_MBH_SMGs,Dietrich2009_Hb_z2,Marziani2009,Bongiorno2014_MM,Banerji2015_redQSOs_hiz,Glikman2015_red_mergers,Suh2015}. 
Likewise, there are several additional $z>5$ quasars with \mgii-based \mbh\ estimates \cite[e.g.,][]{DeRosa2011,DeRosa2014,Wang2015_z5_hiM,Wu2015_z6_nature}.
However, we chose not to include these in our comparative analysis, because of our choice to focus on $z>3$ systems, the small sizes of the samples, and the inhomogeneity the methods of target selection and analysis used in these studies.
We instead focus on the largest samples of unobscured AGNs at $z>3$, selected on the basis of rest-frame UV properties, and for which \mbh\ estimates were derived through an homogeneous spectral analysis.

As \autoref{fig:L_M_Ledd_vs_z} shows, the lower luminosities of the sources studied here are mainly driven by BH masses that are lower than those found for the more luminous \ztpt\ sources analyzed in previous studies, while their accretion rates actually overlap. 
For example, about 85\% of the objects in the combined sample of \cite{Shemmer2004} and \cite{Netzer2007_MBH} have $\mbh>8\times10^{8}\,\Msun$, while about 85\% of the AGNs studied here (save \mysobj) have a mass that is lower than this.
The median \mbh\ of our \ztpt\ AGNs, $\sim5\times10^{8}\,\Msun$, is lower than that of the previously studied sources ($2.4\times10^{9}\,\Msun$) by about 0.7 dex.
On the other hand, the accretion rates of our AGNs -- which span the range $\lledd\sim0.1-0.5$ -- are similar to those found for the more luminous quasars, and also to those of (optically selected) SDSS quasars at $z\sim0.5-1$ \cite[][]{TrakhtNetzer2012_Mg2,Schulze2015_BHMF}.
The obvious outlier in all these comparisons is \mysobj, which has \mbh\ comparable to the most massive SMBHs at $z>2$, and an extremely low accretion rate, of merely $\lledd\simeq0.02$. 
The four \ztpf\ AGNs are powered by yet smaller SMBHs, with typical (median) masses of $\mbh\simeq1.3\times10^{8}\,\Msun$, accreting at normalized rates of $\lledd\simeq0.3$.
These masses are lower, by about an order of magnitude, than those of the faintest AGNs in the combined \ztpf\ sample of \cite{Shemmer2004} and \citet[i.e., those AGNs with $\Lbol \ltsim 3\times10^{46}\,\ergs$]{Netzer2007_MBH}.

\begin{figure}
\centering
\includegraphics[angle=-90, width=0.475\textwidth]{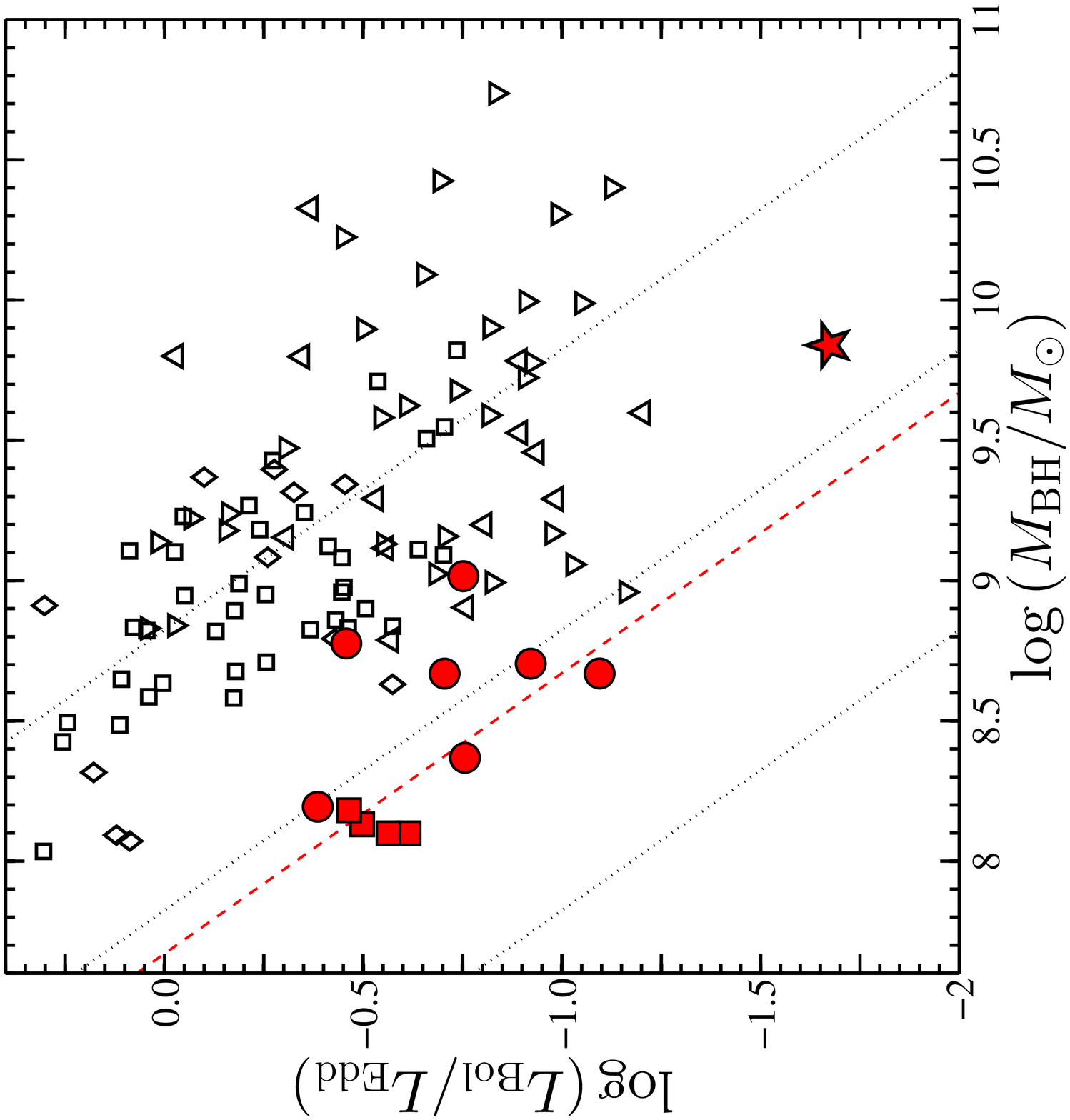} 
\caption{
Accretion rate, in terms of \lledd, vs. black hole mass, \mbh, for the sources studied here and several other relevant samples of high-redshift AGNs.
The symbols are identical to those in \autoref{fig:L_M_Ledd_vs_z}).
The dotted lines represent constant bolometric luminosities of $\Lbol=10^{45}$, $10^{46}$, and $10^{47}\,\ergs$. 
The red dashed line represents the flux limit of our study, $\Lbol=7.8\times10^{45}\,\ergs$ (at $z=3.5$; see \autoref{subsec:sample}), which is most relevant for the \ztpt\ sources. 
Some of the \ztpt\ AGN fall below the flux limit, due to the host-light corrections.
Compared to the combined sample of \citet[][]{Shemmer2004} and \citet[][]{Netzer2007_MBH}, our sources exhibit lower masses but comparable accretion rates.
With the exception of the extreme source \mysobj\ (red star), our sample lacks AGNs with high \mbh\ and low \lledd\ (i.e., $\mbh>2\times10^{9}\,\Msol$ and $\lledd<0.1$).
}
\label{fig:LLedd_vs_MBH}
\end{figure}

As mentioned in \autoref{subsec:sample}, our chosen flux limit for the \ztpt\ AGNs means we could have recovered sources with masses as low as $\mbh\sim7\times10^{7}\,\Msol$ or with accretion rates as low as $\lledd\sim0.01$. 
However, as \autoref{fig:L_M_Ledd_vs_z} demonstrates, the majority of \ztpt\ sources in our sample do not reach these lower limits.
The accretion rates we find ($0.1\ltsim\lledd\ltsim0.5$) are about an order of magnitude above the estimated survey limit.
Given the flux limit of the sample, objects with $\lledd\simeq0.01$ should have $\mbh\simeq5\times10^{9}\,\Msol$ in order to be included in our study.
Indeed, the only object with $\lledd<0.1$ is, again, the extremely massive source \mysobj, which reaches $\lledd\simeq0.02$. 
This low value, as well as other, indirect evidence, indicates that this source is most probably observed at the final stages of SMBH growth, after accreting at much higher rates at yet higher redshifts.
Several previous studies of the distributions of \lledd\ did identify significant populations of intermediate-redshift AGNs ($1<z<2$) with $0.01<\lledd<0.1$ \cite[e.g.,][]{Gavignaud2008_VVDS,Trump2009b_MBH,TrakhtNetzer2012_Mg2,Schulze2015_BHMF}.
Specifically, the low-\lledd\ AGNs studied in \cite{Trump2009b_MBH} and \cite{Schulze2015_BHMF} have BH masses comparable to those studied here.
We conclude that our sample presents compelling evidence for the lack of high-mass, slowly accreting SMBHs - with $\mbh\gtrsim2\times10^{9}\,\Msol$ and $\lledd\ltsim0.1$. 
Such sources would ``fill the gap'' between most of the \ztpt\ sources and \mysobj\ in \autoref{fig:LLedd_vs_MBH}.
However, larger samples are needed to establish this conclusion more firmly.

\subsection{Physical Accretion Rates}
\label{subsec:mdot}

Given reliable estimates of \mbh, and further assuming that the accretion onto the SMBHs occurs within a thin accretion disk, one can derive prescriptions for the estimation of the \emph{physical} accretion rate (i.e., in \mpyr) through the accretion disk, \Maddot.
Several studies derived such prescriptions based on the classical accretion disk model of \cite[e.g., \citealt{Collin2002}]{ShakuraSunyaev1973}, or on more elaborate models that take into account additional complex processes (e.g., general relativistic effects, Comptonization, and winds; see \citealt{DavisLaor2011_AD,NetzerTrakht2014_slim}, and references therein).
Generally, such prescriptions require measurements of the (rest-frame) optical luminosity of the AGNs, which is predominantly emitted by the outer parts of the accretion disk, and is thus mostly unaffected by the spin of the SMBH.

We estimated \Maddot\ for the \Ngood\ AGNs with mass determinations using the prescription presented in \citet[][see also \citealt{DavisLaor2011_AD}]{NetzerTrakht2014_slim}:
\begin{equation}
 \Maddot \simeq 2.4\, \left(\frac{L_{5100,45}}{\cos i}\right)^{3/2}\, \, M_{8}^{-1} \,\,\, \mpyr \,\, ,
 \label{eq:mdot_ad}
\end{equation} 
where $L_{5100,45}\equiv\Lop/10^{45}\,\ergs$, $M_{8}\equiv\mbh/10^{8}\,\Msun$, 
and $\cos i$ represents the inclination of the accretion disk with regard to the line of sight, assumed here to be $\cos i=0.8$ (see \citealt{NetzerTrakht2014_slim} for the full analytical expression and more details).

The resulting accretion rates of the \ztpt\ AGNs are in the range of $\Maddot\sim0.6-6.5\,\mpyr$.
A comparison of the \Maddot\ values obtained through \autoref{eq:mdot_ad} and those estimated from \Lbol\ (\autoref{tab:fit_pars}) suggests that, for most of the sources, the observed data are broadly consistent with a radiatively efficient accretion with $\eta\sim0.1$, as assumed in some of the evolutionary calculations presented in this paper. 
However, we note that a more detailed examination reveals that the \emph{typical} (median) radiative efficiency needed to account for the observed \Lbol, given the \Maddot\ estimates, is somewhat higher, at about $\re\simeq0.15$.
The only outlier is \mysobj\ for which the two \Maddot\ estimates suggest a very high radiative efficiency, reaching (and formally exceeding) the maximum value allowed within the standard accretion disk theory, of $\eta\simeq0.32$.
We note that while \mysobj\ has an extremely low \lledd\ ($\sim$0.02), its physical accretion rate of about $0.4\,\mpyr$ is low but not extreme. 
Two other sources (LID-775 and LID-504) have comparably low \Maddot, despite the fact that their masses are lower than that of \mysobj\ by more than an order of magnitude.
The typically high radiative efficiencies we find are in agreement with the results of several previous studies reporting similar findings for high-mass and/or high-redshift SMBHs, relying either on direct measurements of the iron K$\alpha$ line \cite[][]{Reynolds2014_Ka_spins,Reynolds2014_Einstein_cross}, on arguments similar to the one presented here \cite[e.g.,][]{DavisLaor2011_AD,NetzerTrakht2014_slim,Trakhtenbrot2014_hiz_spin}, or on indirect evidence involving the AGN population as a whole \cite[e.g.,][]{Elvis2002_eta}.

Finally, the \Maddot\ estimates can be used to derive an initial set of estimates of growth time for the SMBHs under study, defined as
$t_{\rm growth, AD}\equiv\mbh/\Mbhdot=\mbh/\Maddot\left(1-\eta\right)$.
Simply assuming $\eta=0.1$, we derive growth times that are generally in the range of $t_{\rm growth, AD}\sim0.1-0.85$ Gyr, again showing that most of the accretion should have happened at higher redshifts. \mysobj\ has an extremely long timescale of $\sim23$ Gyr.
These timescales are generally longer, by a factor of about 1.6, than those derived from \lledd\ alone (see \autoref{subsec:growth} below).


\subsection{Early BH Growth}
\label{subsec:growth}

\begin{figure*}[ht!]
\centering
\includegraphics[width=0.495\textwidth, angle=-90]{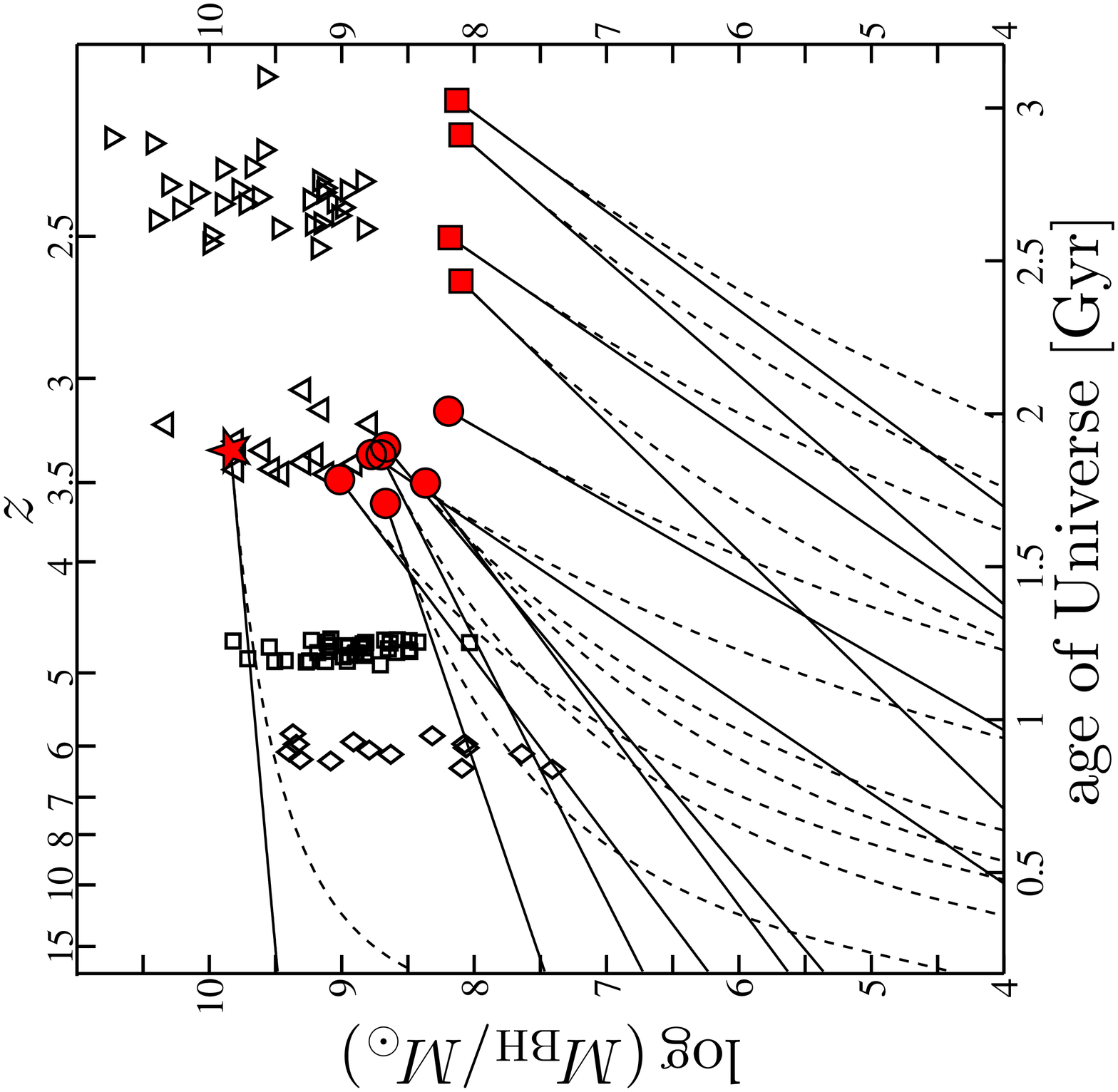}
\hspace{0.015\textwidth}
\includegraphics[width=0.495\textwidth, angle=-90]{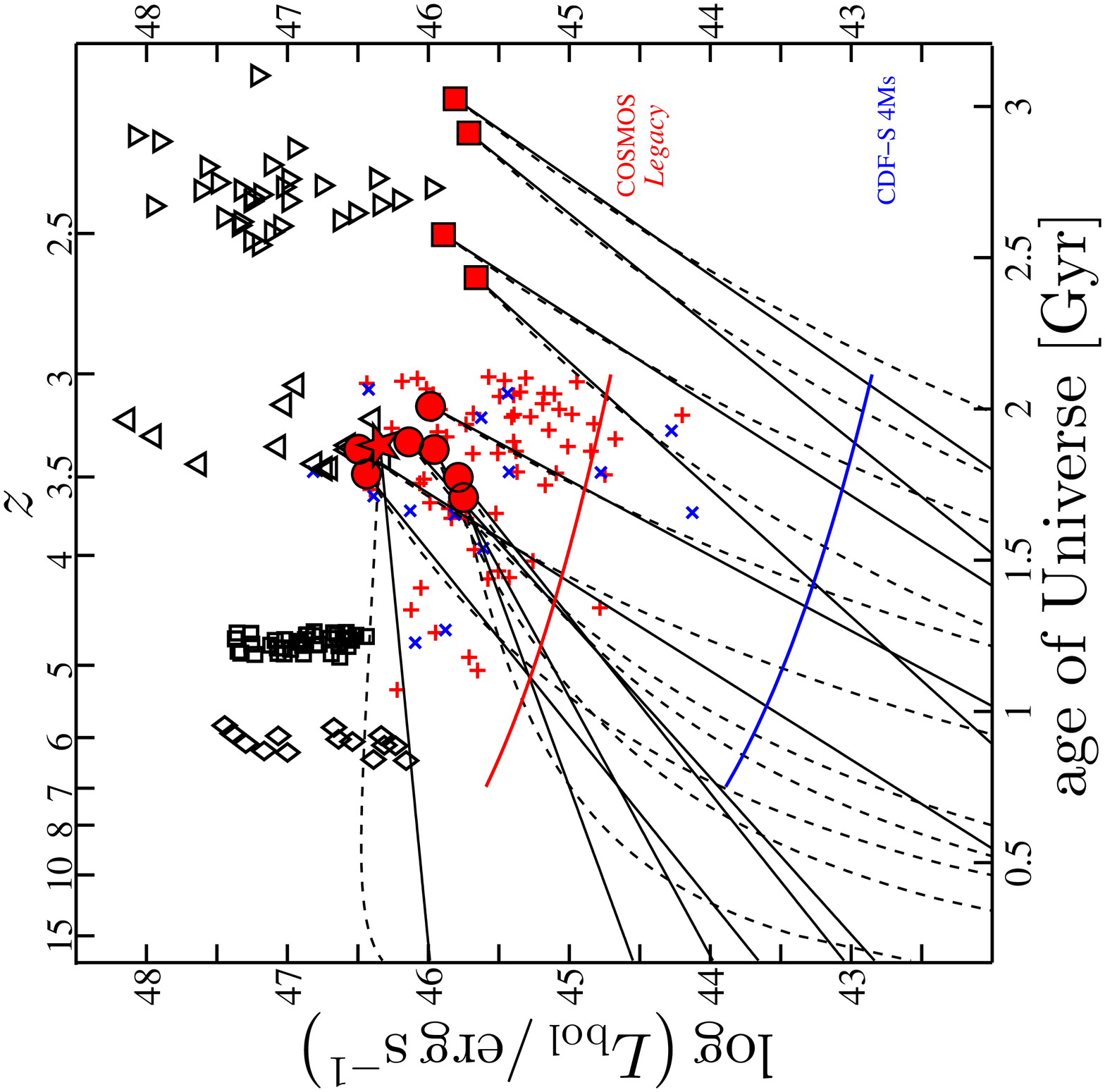} 
\caption{
Calculated evolutionary tracks of \mbh\ and \Lbol\ back to $z=20$, for the sources studied here, compared with other relevant $z>2$ samples (as described in \autoref{fig:L_M_Ledd_vs_z}).
The calculations assume continuous accretion at a (fixed) radiative efficiency of $\eta=0.1$, and accretion rates that are either constant (at observed values), or evolve as $\left(1+z\right)^2$ (illustrated with solid and dashed lines, respectively).
\emph{Left:} evolutionary tracks of \mbh.
Some of the \ztpt\ sources studied here require massive seed BHs, with $\mseed \gtrsim 10^{4}\,\Msun$, and/or a higher accretion rate in previous epochs.
For the extreme source \mysobj, these calculation strongly support a scenario in which the SMBH used to accrete at \emph{much} higher rates at $z\gtrsim3.5$.
The \ztpf\ sources can be easily explained by stellar BH seeds, even if invoking a non-unity duty cycle.
\emph{Right:} evolutionary tracks of \Lbol. 
Here we also plot high-$z$ X-ray-selected samples with spectroscopic redshifts from the \emph{Chandra COSMOS Legacy} (red ``+''; M15) and the 4Ms CDF-S \cite[blue ``$\times$'';][]{Vito2013_CDFs_hiz} surveys.
The flux limits of these surveys are indicated as colored dashed lines (assuming the \citealt{Marconi2004} bolometric corrections).
Both surveys should, in principle, detect the progenitors of our sample of AGNs, up to $z\sim5-6$. 
However, such faint AGNs are detected at very small numbers, if at all (see discussion in text).
}
\label{fig:M_vs_t_back}
\end{figure*} 



Assuming a SMBH accretes matter with a constant \lledd\ and radiative efficiency \re, its mass increases exponentially with time, with a typical $e$-folding timescale of
\begin{equation}
\tau = 4 \times 10^8 ~~ \frac {\eta /(1- \eta) }{\lledd} ~ {\rm yr} . 
\label{eq:tau_salpeter}
\end{equation}
If one further assumes a certain initial (seed) BH mass, \mseed, then the time required to grow from \mseed\ to the observed \mbh, \tgrow, is
\begin{equation}
  \tgrow = \tau ~ \ln
  \left( \frac{\mbh}{\mseed} \right) ~ {\rm yr} .
\label{eq:t_grow}
\end{equation}
For the \ztpt\ AGNs studied here, the $e$-folding timescales are in the range $0.1-2$ Gyr, assuming $\eta=0.1$. 
For the lower-redshift sources the timescales are shorter, at about 0.1 Gyr.
Further assuming that $\mseed=100$, $10^{4}$, or $10^{6}\,\Msun$ results in growth times in the range of $1.5-8.5$, $1-6$, or $0.5-3.4$ Gyr, respectively, for the \ztpt\ sources, excluding \mysobj.
The atypically low accretion rate of \mysobj\ translates to an $e$-folding timescale of 2 Gyr. Even in the most favorable scenario of $\mseed=10^6\,\Msol$, the growth time is longer than the age of the Universe (at the observed epoch), suggesting that \mysobj\ must have experienced a dramatic drop in \lledd\ (see T15 for a detailed discussion).

In \autoref{fig:M_vs_t_back} we illustrate several evolutionary tracks for the SMBHs in our sample, since $z=20$.
The simplest scenario assumes that each SMBH grows with a constant \lledd, fixed to the observed value.
The points where each of the (diagonal solid) lines crosses the y-axis of the \emph{left} panel of \autoref{fig:M_vs_t_back} may be considered as the implied (seed) BH mass at $z=20$, under these assumptions.
The \ztpf\ sources have high-enough accretion rates to account for their observed masses, even if one assumes that they originate from ``stellar'' BH seeds ($\mseed\lesssim100\,\Msun$) and/or a fractional duty cycle for accretion.
Among the \ztpt\ sources, however, we see some evidence for either more massive seeds and/or higher accretion rates in yet earlier epochs, as the implied seed masses are typically of order $\mseed\sim10^5\,\Msun$.
To illustrate the effect of having higher \lledd\ at earlier epochs, we repeated the calculation of evolutionary tracks, this time assuming that \lledd\ increases with redshift, as suggested by several studies of higher-luminosity AGNs (see \autoref{fig:L_M_Ledd_vs_z}, and also \citealt{DeRosa2014}).
We assume two very simple evolutionary trends, of the form $\lledd\propto\left(1+z\right)$ and $\lledd\propto\left(1+z\right)^2$, both capped at the Eddington limit (i.e., $\lledd\leq1$). 
The stronger evolutionary trend is consistent with a fit to all the data points in the bottom panel of \autoref{fig:L_M_Ledd_vs_z}.
The results of this latter calculation are illustrated as dashed lines in \autoref{fig:M_vs_t_back}.
\footnote{As for the maximal allowed \lledd, we note that few of the \zsix\ and \zfpe\ sources have observed accretion rates above the Eddington limit, but those could well be due to the uncertainties related to \lledd\ estimation.}
These calculations suggest that massive seeds are required to explain \emph{some} \ztpt\ sources, even under these favorable conditions.
The only scenario in which all the implied seed masses are in the ``stellar'' regime is indeed the one with the strongest evolution in accretion rates, $\lledd\propto\left(1+z\right)^2$.
We note, however, that all these calculations assume continuous growth, i.e. a duty cycle of 100\%. 
Any other, more realistic choice for the duty cycle, as well as the indirect evidence for somewhat elevated radiative efficiencies for some of the AGNs (\autoref{subsec:mdot}), would further challenge the ability of stellar BH seeds to account for the observed BH masses.

Another interesting point that is clearly evident from \autoref{fig:M_vs_t_back} is that most of the SMBHs studied here \emph{cannot} be considered as the descendants of the known higher-redshift SMBHs.
This is due to the simple fact that the observed masses of the \zpaper\ SMBHs are lower than, or comparable to, those of the higher-redshift ones.
The only exception for this interpretation (except for \mysobj) would be a scenario where the lowest-mass SMBHs at \zsix\ would shut off their accretion within a very short timescale, and then be briefly ``re-activated'' at $z\sim3.5$. 
However, given the large difference between the number densities of the population from which our sample is drawn and that of the higher-redshift, higher-luminosity samples shown in \autoref{fig:M_vs_t_back} \cite[e.g.,][]{McGreer2013_QLF_z5}, this scenario is unlikely.

The evolutionary tracks we calculate for our \ztpt\ sources, combined with the associated number density of their parent population, strongly support the existence of a significant population of relatively low-mass ($\mbh\sim10^{6-7}\,\Msun$), active SMBHs at $z\sim5-7$. 
Moreover, as the right panel of \autoref{fig:M_vs_t_back} shows, such sources should be observable, as their luminosities are expected to exceed the flux limits of existing deep X-ray surveys, such as the \XVP\ itself, or the 4 Ms CDF-S survey \cite[][]{Xue2011_CDFS_4Ms}.
However, very few such sources are indeed detected.
Several surveys of optically selected, unobscured AGNs at $z\sim5-7$ suggest number densities of order $10^{-8}\,\NDunit$ \cite[e.g.,][and references therein]{McGreer2013_QLF_z5,Kashikawa2015_QLF}.
Even when combining all currently available X-ray surveys, and including \emph{all} sources with redshifts $z\sim5$, the number density of the sources that have comparable luminosities to what we predict here ($\log\Lhard\sim43-43.5$) is roughly $\sim5\times10^{-7}\,\NDunit$.
In particular, the recent study of \cite{Marchesi2015_XVP_hiz} identified about 30 X-ray AGNs at $z>4$, based on the same X-ray \chandra\ data used for the selection of the sample studied here.
Of these sources, nine are at $z>5$ and only four are at $z\geq6$, with the vast majority of such high-$z$ sources having only photometric redshift estimates.
In terms of the typical luminosities of these AGNs, the right panel of \autoref{fig:M_vs_t_back} clearly shows that the $z\sim5$ X-ray AGNs can indeed be considered as the parent population of our sources.
However, the number density of such high-$z$ AGNs is significantly lower than that of our sample.
The Marchesi et al. study shows that the cumulative number density of X-ray-selected AGNs drops dramatically with increasing redshift, to reach $\Phi\sim5\times10^{-7}\,\NDunit$ by $z\simeq5$ (split roughly equally between obscured and unobscured AGNs), and to about $10^{-7}\,\NDunit$ by $z\sim6$.
This is about an order of magnitude lower than what we consider for the $z\sim6$ progenitors of our sources.
This discrepancy is \emph{not} driven by the (X-ray) flux limit of the \XVP.
Indeed, the study of \cite{Weigel2015_CDFS} did not identify any (X-ray-selected) $z\gtrsim5$ AGNs in the 4 Ms CDF-S data, the deepest available survey \cite[][]{Xue2011_CDFS_4Ms}.\footnote{Another recent study by \cite{Giallongo2015_CANDELS_hiz_AGN} did claim to identify several $z>4$ sources in the CDF-S field. However, their technique for identifying X-ray sources  goes far beyond the standard procedures used in the X-ray luminosity function studies we refer to here, and may introduce false detections.}
As illustrated in the right panel of \autoref{fig:M_vs_t_back}, the 4 Ms CDF-S data should have easily detected the progenitors of our sources.
We note that the lack of such higher-redshift sources is \emph{not} due to the small size of the CDF-S survey, because it does contain some high-luminosity AGNs at $z\sim5$. 
In principle, given the general behavior of luminosity functions, the lower-luminosity progenitors of our \ztpt\ AGNs should have been even more numerous.
We conclude that our sample provides compelling evidence for the existence of a significant population ($\Phi\sim10^{-6}\,\NDunit$) of faint $z\sim5-6$ AGNs, powered by SMBHs with $\mbh\sim10^{6-7}\,\Msol$ and $\Lbol\sim\left(1-3\right)\times10^{44}\,\ergs$, which, however, is not detected (at sufficiently large numbers) in the currently available deep X-rays surveys. 
We note that while the decline in the number density of AGNs at $z>3$ was well established in several previous studies, including those based on \chandra\ data in COSMOS \cite[][M15]{Civano2011_hiz}, our analysis clearly demonstrates that such ``progenitor'' AGNs are \emph{expected}, given the masses and accretion rates of the \ztpt\ AGNs.

There are several possible explanations for this apparent discrepancy between the expected and observed number of $z\gtrsim5$ AGNss:
\renewcommand{\labelenumi}{\roman{enumi}.}
\begin{enumerate}
 \item 
First, the small number of detected ``progenitor'' systems can be explained by a high fraction of obscured AGNs (\fobs).
If the obscuration of each accreting SMBH evolves with the luminosity of the central source, then we should expect that a certain fraction of the progenitors of our sources would be obscured at earlier epochs. Such a scenario is expected within the framework of ``receding torus'' models \cite[e.g.,][]{Lawrence1991_receding}, where lower luminosities are typically associated with a higher \fobs.
However, several recent studies show that there is little observational evidence in support of such torus models (see, e.g., \citealt{Oh2015_hidden_BLAGN,Netzer2016_Herschel_hiz}, and \citealt{Netzer2015_torus_rev} for a recent review).
There is, however, somewhat stronger evidence for an increase in the typical \fobs\ toward high redshifts \cite[e.g.,][]{Treister2006_obs_evo,Hasinger2008}, perhaps in concert with an increasing frequency of major galaxy mergers \cite[e.g.,][]{Treister2010_fobs_z}.
A more plausible scenario is therefore that the progenitors of our \ztpt\ AGNs are embedded in dusty galaxy merger environments with high-column density. 

\item
Second, it is possible that, early on, our sources grew with lower radiative efficiencies, which would result in yet-lower luminosities per given (physical) accretion rate. 
To illustrate the possible effects of lower \re\ on the projected evolutionary tracks of our sources, we repeated the aforementioned evolutionary calculations with $\re=0.05$ (comparable to the lowest possible value within the standard model of a thin accretion disk). 
Indeed, at $z\gtrsim5$ the expected luminosities are significantly lower than those projected under the fiducial assumptions.
The differences amount to at least an order of magnitude at $z\sim5$, and at least a factor of 30 at $z\sim6$, making most of these projected progenitors undetectable even in the deepest surveys.
In this context, we recall that the efficiencies we infer for the sources are actually somewhat \emph{higher} than standard ($\eta\simeq0.15$; \autoref{subsec:mdot}). 
However, lower efficiencies at earlier times may still be expected if one assumes, for example, a relatively prolonged accretion episode that (gradually) ``spins up'' the SMBHs \cite[e.g.,][and references therein]{Dotti2013} or supercritical accretion through ``slim'' accretion disks \cite[e.g.,][]{Madau2014_supEdd}.

\item
Finally, the discrepancy may be explained in terms of the AGN duty cycle, on either long (host-scale fueling) or short (accretion flow variability) timescales.
In the present context, this would require that high-redshift, lower-luminosity AGNs would have a lower duty cycle than their (slightly) lower-redshift descendants. 
We note that such a scenario would actually further complicate the situation, as the growth of the SMBHs would be slower. 
This, in turn, would mean that our sources should be associated with progenitors of yet higher luminosity at $z\gtrsim5$, which have yet lower number densities.
\end{enumerate}
\renewcommand{\labelenumi}{\arabic{enumi}.}

\noindent
We conclude that the simplest explanation for the discrepancy between the observed and expected properties of the progenitors of our \ztpt\ AGNs is probably due to a combination of an evolution in  the radiative efficiencies and/or obscuration fractions, during the growth of individual systems. 
We stress that such trends are beyond the scope of most ``synthesis models,'' which assume time-invariable accretion rates, radiative efficiencies, and/or obscuration fractions \cite[e.g.,][and references therein]{Ueda2014,Georgakakis2015_XLF_hiz}.

\vspace{0.2cm}

\section{Summary and Conclusion}
\label{sec:summary}

We have presented new Keck/MOSFIRE \kband\ spectra for a total of \Ntot\ unobscured, $z\sim2.1-3.7$ AGNs, selected through the extensive \chandra\ X-ray coverage of the COSMOS field.
We mainly focus on \Nhb\ objects at \ztpt, representing a parent population with a number density of roughly $10^{-6}-10^{-5}\,\NDunit$ - a factor of $\sim25$ more abundant than previously studied samples of AGNs at these high redshifts.
The new data enabled us to measure the black hole masses (\mbh) and accretion rates (both in terms of \lledd\ and \Maddot) for these sources, and to trace their early growth.
Our main findings are as follows:
\begin{enumerate}
\setlength{\itemsep}{7pt plus 0pt minus 0pt}

\item 
The \ztpt\ AGNs are powered by SMBHs with typical masses of $\mbh\sim5\times10^8\,\Msol$ and accretion rates of $\lledd\sim0.1-0.4$. 
These BH masses are significantly lower than those found for higher-luminosity AGNs at comparable redshifts.
Our sample generally lacks AGNs powered by high-mass but slowly accreting SMBHs (i.e., $\lledd<0.1$), although such systems are well within our chosen flux limit.
Assuming a standard, thin accretion disk, the data suggest somewhat higher-than-typical radiative efficiencies, of about $\re\sim0.15$, in agreement with several recent studies.
 
\item 
Assuming continuous growth at the observed accretion rates, most of the \ztpt\ SMBHs had to grow from massive BH seeds (i.e., $\mseed>10^4\,\Msol$). 
Stellar seeds can only account for the observed masses if \lledd\ was higher at yet earlier epochs. 
However, invoking any reasonable duty cycle for the accretion, as well as the indirect evidence for somewhat higher-than-standard radiative efficiencies, further complicates the scenario of stellar BH seeds. 
 
\item Our analysis predicts the existence of a large population of $z\sim6-7$ AGNs, with $\Phi\sim10^{-5}\,\NDunit$, $\mbh\sim10^6\,\Msol$, and $\Lhard\gtrsim10^{43}\,\ergs$. Such sources are not detected in sufficiently large numbers in the existing deep X-ray surveys, perhaps because of increased obscuration at high redshift and/or because of lower radiative efficiences in the early stages of black hole growth.

\item Two of the \ztpt\ sources, and possibly one additional source ($\sim17-25\%$) have extremely weak broad \hbeta\ emission components, although their (archival) optical spectra clearly show strong emission from other, high-ionization broad lines (e.g., \civ). 
The weakness of the broad \hbeta\ lines \emph{cannot} be due to dust obscuration along the line of sight, nor due to the lack of BLR gas. 
A sudden decrease in AGN (continuum) luminosity is also improbable.
Another source shows a peculiarly broad \oiii\ profile.
Repeated optical spectroscopy of these sources may clarify the physical mechanisms that drive the highly unusual broad-line emission.

\item One source in our sample, the broad-absorption-line AGN \mysobj, has a significantly higher \mbh\ and lower \lledd\ than the rest of the sample.
Our detailed analysis \cite[published separately as][]{Trakhtenbrot2015_CID947} suggests that the SMBH in this system is at the final phase of growth.
Compared with the rest of the sample analyzed here, \mysobj\ appears to be an outlier in the general distributions of \mbh\ and \lledd. 
We stress, however, that it is highly unlikely that systems like \mysobj\ are extremely rare, as we have identified one such object among a sample of ten.
 
\end{enumerate}

Our sample presents preliminary insights into key properties of \emph{typical} SMBHs at \ztpt.
Clearly, a larger sample of faint AGNs is needed in order to establish the black hole mass function and accretion rate function at this early cosmic epoch. 
We are pursuing these goals by relying on the (relatively) unbiased selection function enabled by deep X-ray surveys, in extragalactic fields where a rich collection of supporting multi-wavelength data are available.
A forthcoming publication will explore the host galaxies of the AGNs studied here, and trace the evolution of the well-known SMBH-host scaling relations to $z\sim3.5$.

\acknowledgements
We thank the anonymous referee, whose numerous suggestions helped improve the paper.
The new MOSFIRE data presented here were obtained at the W.\ M.\ Keck Observatory, which is operated as a scientific partnership among the California Institute of Technology, the University of California, and the National Aeronautics and Space Administration. 
The Observatory was made possible by the generous financial support of the W.\ M.\ Keck Foundation. 
We are grateful for the support from Yale University that allows access to the Keck telescopes.
We thank M.\ Kassis, L.\ Rizzi, and the rest of the staff at the W.\ M.\ Keck observatories at Waimea, HI, for their support during the observing runs.
We recognize and acknowledge the very significant cultural role and reverence that the summit of Mauna Kea has always had within the indigenous Hawaiian community.  
We are most fortunate to have the opportunity to conduct observations from this mountain.
Some of the analysis presented here is based on data products from observations made with European Southern Observatory (ESO) Telescopes at the La Silla Paranal Observatory under ESO program ID 179.A-2005 and on data products produced by TERAPIX and the Cambridge Astronomy Survey Unit on behalf of the UltraVISTA consortium.
This work made use of the MATLAB package for astronomy and astrophysics \cite[][]{Ofek2014_matlab}.
We thank A.\ Weigel and N.\ Caplar for beneficial discussions.
F.C. and C.M.U. gratefully thank Debra Fine for her support of women in science. 
This work was supported in part by NASA \chandra\ grant numbers  GO3-14150C and GO3-14150B (F.C., S.M., H.S., M.E.).
K.S. gratefully acknowledges support from Swiss National Science Foundation Grant PP00P2\_138979/1. 
J.M. acknowledges support for his PhD by CONICYT-PCHA/doctorado Nacional para extranjeros, scholarship 2013-63130316.
A.F. acknowledges support from the Swiss National Science Foundation. 




\end{document}